\documentclass[a4paper,m11pt]{article}
\usepackage{jheppub} 
\usepackage{lineno}
\usepackage{graphicx}
\usepackage{lipsum}
\usepackage{slashed}
\usepackage{mathtools}
\usepackage{bm}
\usepackage{tikz-feynman, contour}
\usepackage{tikz-cd}
\usepackage{bbm}
\usepackage{subcaption}
\usepackage{soul}
\usepackage{float}
\restylefloat{table}
\usepackage{xcolor}
\usepackage[normalem]{ulem}
\usepackage{array}
\usepackage{tensor}
\usepackage{makecell}
\allowdisplaybreaks

\usepackage{xspace}

\newcommand{\be}{\begin{equation}}
\newcommand{\ee}{\end{equation}}
\newcommand{\bea}{\begin{eqnarray}}
\newcommand{\eea}{\end{eqnarray}}
\newcommand{\beq}{\begin{equation}}
\newcommand{\eeq}{\end{equation}}

\newenvironment{nalign}{ 
	\begin{equation}
	\begin{aligned}
}{
	\end{aligned}
	\end{equation}
	\ignorespacesafterend
}

\begin{document}
\title{Covariance of Scattering Amplitudes from Counting Carefully}

\author[a]{Mohammad Alminawi}

\affiliation[a]{Physik Institut, Universit\"at Z\"urich, Switzerland}
\emailAdd{mohammad.alminawi@physik.uzh.ch}

\abstract{Invariance of on-shell scattering amplitudes under field redefinitions is a well known property in field theory that corresponds to covariance of on-shell amputated connected functions. In recent years there have been great efforts to define a formalism in which the covariance is manifest at all stages of calculation, mainly resorting to geometrical interpretations. In this work covariance is analysed using combinatorial methods relying only on the properties of the tree level effective action, without referring to specific formulations of the Lagrangian. We provide an explicit proof of covariance of on-shell connected functions and of the existence of covariant Feynman rules and we derive an explicitly covariant closed formula for tree level on-shell connected functions with any number of external legs.}

\maketitle
\section{Introduction}
The invariance of physical amplitudes under certain transformations known as {\em field redefinitions} is a classical result in quantum field theory. When we compute an observable within the path integral formalism, such as a scattering amplitude, the `field' $\phi(x)$ may be treated as an integration variable. The invariance of observables under field redefinitions is no different than the invariance of an integral under a change of variables \cite{Arzt1995:Covariance,CHISHOLM1961:Fieldredefinitions,KAMEFUCHI1961:FieldRedefinitions}. 

Scattering amplitudes are obtained from $\mathcal{S}$-matrix elements using the Lehmann–Symanzik
–Zimmermann (LSZ) formula ~\cite{LSZ:1955}
\begin{equation}
    \left\langle p_n \ldots p_{m+1}\right| i \mathcal{T}\left|p_1 \ldots p_m\right\rangle_{A_1 \ldots A_n}=(2 \pi)^4 \delta^{(4)}\left(\sum p_i\right) \bigg[\prod_{i=1}^n e^{a_i}_{A_i}\bigg]\mathcal{A}_{a_1 \ldots a_n}
\end{equation}
where $\mathcal{S} = 1 - i \mathcal{T}$, $\bigg[\prod_{i=1}^n e^{a_i}_{A_i}\bigg]$ is a projection on the mass eigenstates $A_i$ and $\mathcal{A}_{a_1 \dots a_n}$ denotes the amputated connected function for $n$ external states. As seen from the LSZ formula, the invariance of scattering amplitudes under field redefinitions is equivalent to the covariance of the amputated connected functions. 

Despite the apparent simplicity of showcasing covariance and invariance of amplitudes by appealing to properties of path integrals, the infinite-dimensional nature of the integrals typically means such arguments are not rigorous, and also limits the usability of such results. In practice, quantum field theories (QFTs) are typically defined perturbatively through the use of Lagrangian densities, which are not invariant under field redefinitions. Two seemingly different Lagrangians may in fact correspond to the same physical theory, in the sense that they produce identical physical observables, due to the two Lagrangians being related by a field redefinition.

To resolve this well known ambiguity, it is valuable to develop computational techniques that are manifestly covariant, and to try to understand the structure of Lagrangians in terms of covariant objects. In the context of physics Beyond the Standard Model (BSM), the use of effective field theories (EFTs) makes the issue particularly pertinent, as the BSM effects are captured via either of the Standard Model Effective Field Theory (SMEFT) or the Higgs Effective Field Theory (HEFT), which differ from one another in how the scalar sector is parametrized. A SMEFT Lagrangian may always be recast into a HEFT Lagrangian, but the inverse map is singular, meaning there exist BSM scenarios that can match only to HEFT, not SMEFT~\cite{Cohen:2020xca,Falkowski2019:WhichEFT,Alonso_2016:GeometryScalar,Alonso_2015:GeometryHiggs, Helset_2020:GeoSmeft}. This presents a need to study the physical amplitudes or equivalently the connected amputated functions $\mathcal{A}_{a_1 \dots a_n}$ for arbitrary values of $n$ to identify whether a HEFT theory encodes the same physics as a SMEFT theory. 

While covariance is a fundamental property of $\mathcal{A}_{a_1 \dots a_n}$, it can be explicitly highlighted through the use of different formalisms, with the use of geometry to represent the Lagrangian emerging as a favored approach for distinguishing HEFT theories from SMEFT theories in recent years; by identifying the fields with maps into a manifold, the kinetic (two derivative) term naturally finds a description in terms of a (pseudo-)Riemannian metric $g_{ij}(\phi)\partial_\mu \phi^i \partial^\mu \phi^j$, which in turn allows for its contributions to the amputated connected function to be expressed in terms of covariant geometric objects ~\cite{Alonso_2015:GeometryHiggs,Alonso_2016:GeometryScalar,Alonso_2016:SigmaModels,Cohen:2020xca}. The Levi-Civita connection defined by the metric tensor makes the derivation of covariant Feynman rules in geometric theories particularly simple~\cite{Apostolos2020:Covariance, alminawi2025scalaramps}.  

The computation of physical observables is typically carried out using perturbation theory, which can be rigorously encoded via the usual diagrammatic approach, wherein the $\mathcal{A}_{a_1 \dots a_n}$ are pieced together by gluing one particle irreducible (1PI) functions that are obtained by taking functional derivatives of an effective action $\Gamma[\phi]$, the leading order contributions come from tree diagrams, which will be the main interest of this paper.

In more detail, in the diagrammatic approach the amputated connected functions are obtained via Wick contractions of 1PI vertices with propagators. The 1PI functions themselves are not individually covariant in general. Consequently, in the diagrammatic approach  the covariance of connected functions is obscured behind a series of cancellations that must occur between the different Feynman diagrams. This structure has been examined  across different contexts, using tools from combinatorics in ~\cite{FIGUEROA:HopfAlgebra,Kreimer:Combinatorics2016,Balduf:Combinatorics2020}.
Restoring covariance at the level of 1PI functions was demonstrated using the path integral formalism in ~\cite{Apostolos2020:Covariance,HONERKAMP1972:Multiloops}, with the validity being showcased through computations in specific examples relevant to pion physics in  ~\cite{HONERKAMP1974:OnShell, ECKER1972:CovariantPerturbation}.

Computing $\mathcal{A}_{a_1 \dots a_n}$ for high values of $n$, which is a challenging task, has been tackled by several previous works, which this paper will draw upon. Combinatorial approaches have provided an iterative result for the "tree sums" ~\cite{Kreimer:Combinatorics2016,Balduf:Combinatorics2020}. However, the result does not distinguish the different possible index contractions among identical vertices. The path integral formalism has been used to derive a recursive method for the computation of Feynman rules~\cite{Cohen:2023Covariance,Cohen:2024Geometry,cohen2025geometricbuildingblocks}, which produce the amputated functions $\mathcal{A}_{a_1 \dots a_n}$ up to "evanescent" terms, i.e terms that vanish on-shell. The formalism has also been used to derive manifestly covariant Feynman rules~\cite{Apostolos2020:Covariance,HONERKAMP1972:Multiloops}.

In this work, previous works are expanded upon and connected to one another, to provide a more explicit understanding of the covariance of $\mathcal{A}_{a_1 \dots a_n}$ under field redefinitions as well as exploiting it to develop an efficient method for computing the connected function for arbitrary $n$. In more detail:

\begin{itemize}
    \item An alternative combinatorial approach is adopted, based solely on the effective action $\Gamma[\phi]$ in combination with the vacuum and on-shell conditions. No assumptions are made regarding the type of fields involved nor the structure of the interaction terms, thus deriving the properties of the $\mathcal{S}$-matrix elements directly from the effective action, hence it holds for a generic EFT.\\
    \item A refined counting function is derived using generating functionals, which correctly identifies the dimension of the automorphism group of any tree level Feynman diagrams at arbitrary $n$, thus accounting for all possible index contractions.\\
    \item The refined counting function and the Fa\`a di Bruno formula are used to examine the transformation behaviors of all relevant tree diagrams proving the covariance of amputated connected functions for arbitrary $n$.\\
    \item The existence of covariant Feynman rules and the validity of using them to compute amputated connected functions is shown for any $n$ at the tree level, the efficiency of this approach is demonstrated in \cite{alminawi2025scalaramps} using covariant Feynman rules up to $n=10$. The validity of the results for an arbitrary number of derivatives also makes them a valuable asset for the application of geometric techniques to higher derivatives through the use of jet bundles \cite{Alminawi_2024:JetBundles} or functional geometry ~\cite{Apostolos2020:Covariance,cohen2025geometricbuildingblocks, Helset2022:Geometry, Cohen:2024Geometry, Cheung2022:GeometryKinematics,Cohen:2023Covariance}
\end{itemize}

The extension of the counting and computational methods to loops is the natural next step. A naive generalization by replacing the tree level effective action everywhere with a higher order one can give a valuable hint on how to address the issue, but to truncate the result to $l$ loops an additional expansion of the perturbative series is needed, which will be explored in a future work.

The structure of the paper is as follows: the covariance of the amputated $n$-point connected function is introduced as a combinatoric problem in Sec. ~\ref{Sec2:Covariance as a combinatoric property}. Feynman diagrams are closely tied to graph theory and combinatoric concepts, the process of translating notation is handled in Sec. ~\ref{sec:Counting Tree Graphs}, first translating the problem into integer partitions in ~\ref{subsec: Trees as Integer Partitions} and then converting tree level Feynman diagrams into tree graphs in ~\ref{subsec: Feynman diagrams to graphs}, the remaining two subsections are dedicated to counting topologies, with the method of generating functions being introduced and used to derive the refined counting in ~\ref{subsec: Refined counting}. Covariance is examined in Sec ~\ref{sec: On-Shell Connected Functions and Covariant Feynman Rules}, first by proving it for a general case by imposing the physical conditions and using the Fa\`a di Bruno formula in ~\ref{subsec: Proof of covariance}, then using the same approach to prove the existence and validity of covariant Feynman rules to arbitrary $n$ in ~\ref{subsec: Covariant Feynman rules}. The paper concludes in Sec ~\ref{sec: Conclusion} where the new results are reviewed and possible follow ups are mentioned.

\section{Covariance as a combinatoric property}
\label{Sec2:Covariance as a combinatoric property}

The covariance of the connected amputated $n$-point function on-shell, particularly at the tree level, can easily be showcased through the use of generating functionals as done in~\cite{Cohen:2023Covariance,Cohen2023:Letter}. However, in practice the derivation of $\mathcal{A}_{a_1\dots a_n}$ often follows a different path, starting from the effective action $\Gamma[\phi]$, we can identify 1PI functions as the functional derivatives of the effective action
\begin{equation}
    F_{a_1 \dots a_n} = \frac{i\delta^n \Gamma }{\delta \phi^{a_1} \dots \delta \phi^{a_n}}\bigg|_{\phi = \phi_\text{cl}}
\end{equation}
where $\phi_\text{cl} = \langle 0|\phi|0 \rangle$ is the classical field configuration defined by the vacuum condition $\delta \Gamma/\delta \phi |_{\phi = \phi_\text{cl}} = 0$.

The 1PI functions are not manifestly covariant under a transformation $\phi \to \psi(\phi)$ for arbitrary $n$ with the special exceptions of $n=1$ defining the vacuum condition
\begin{equation}
    0=\frac{i\delta \Gamma[\psi(\phi)] }{\delta \phi^{a_1}}\bigg|_{\phi = \phi_\text{cl}} =  \frac{\partial \psi^{b_1}}{\partial \phi^{a_1}}\frac{i\delta \Gamma[\psi(\phi)] }{\delta \psi^{b_1}}\bigg|_{\phi = \phi_\text{cl}}
\end{equation}
and $n=2$ which gives
\begin{nalign}
    \frac{i\delta^2 \Gamma[\psi(\phi)] }{\delta \phi^{a_1}\delta \phi^{a_2}}\bigg|_{\phi = \phi_\text{cl}} &=  \frac{\partial \psi^{b_1}}{\partial \phi^{a_1}} \frac{\partial \psi^{b_2}}{\partial \phi^{a_2}}\frac{i\delta^2 \Gamma[\psi(\phi)] }{\delta \psi^{b_1} \delta \psi^{b_2}}\bigg|_{\phi = \phi_\text{cl}}+  \frac{\partial^2 \psi^{b_1}}{\partial \phi^{a_1} \partial \phi^{a_2}}\frac{i\delta \Gamma[\psi(\phi)] }{\delta \psi^{b_1}}\bigg|_{\phi = \phi_\text{cl}} \\&=  \frac{\partial \psi^{b_1}}{\partial \phi^{a_1}} \frac{\partial \psi^{b_2}}{\partial \phi^{a_2}}\frac{i\delta^2 \Gamma[\psi(\phi)] }{\delta \psi^{b_1} \delta \psi^{b_2}}\bigg|_{\phi = \phi_\text{cl}}
\end{nalign}
the $n=2$ case defines the inverse propagator $D_F^{-1}$, which as showcased above is covariant once the vacuum condition is imposed. The on-shell condition for mass eigenstates can be defined using the 2-point 1PI function, it reads \footnote{The eigenvalues of the 2 point function vanish, rather than the full matrix. However, it is possible to choose a diagonal basis, as permitted field redefinitions do not change the eigenvalues}
\begin{equation}
    \frac{i\delta^2 \Gamma[\psi] }{\delta \psi^{b_1} \delta \psi^{b_2}} \bigg|_{\psi = \psi_\text{cl},\text{on-shell}} = 0
\end{equation}

For an arbitrary $n$ the transformation behavior of the 1PI function is encoded by the Fa\`a di Bruno formula~\cite{Comtet_Combinatorics, Riordan_Combinatorial}
\begin{equation}
    \frac{\delta^n}{\delta \phi^n} \Gamma[\psi(\phi)] = \sum_{k=1}^n  \frac{\delta^k}{\delta \psi^k} \Gamma[\psi(\phi)] B_{n, k}\left(\psi^{\prime}(\phi), \psi^{\prime \prime}(\phi), \ldots, \psi^{(n-k+1)}(\phi)\right)
\end{equation}
where $\psi' = \frac{\delta \psi}{\delta \phi}$ and $B_{n,k}$ are the partial exponential Bell polynomials defined by
\begin{equation}
B_{n, k}\left(x_1, x_2, \ldots, x_{n-k+1}\right)=\sum \frac{n!}{j_{1}!j_{2}!\cdots j_{n-k+1}!}\left(\frac{x_1}{1!}\right)^{j_1}\left(\frac{x_2}{2!}\right)^{j_2} \cdots\left(\frac{x_{n-k+1}}{(n-k+1)!}\right)^{j_{n-k+1}}
\end{equation}
the polynomials denote the number of ways that a set of $n$ distinct elements can be partitioned into $k$ subsets. 

Under an arbitrary field redefinition $\phi \to \phi + f(\phi)$ the new Feynman rules emerging will be dependent on terms from Feynman rules with fewer legs, for example the four point Feynman rule with suppressed indices after the transformation is given by
\begin{equation}
  \frac{\delta^4}{\delta \phi^4} \Gamma[\psi(\phi)] = \frac{\delta^4 \Gamma[\psi] }{\delta \psi^4} \psi'^4 + 6\frac{\delta^3 \Gamma[\psi]}{\delta \psi^3}  \psi'^2 \psi'' + 3 \frac{\delta^2 \Gamma[\psi]}{\delta \psi^2} \psi''^2 + 4 \frac{\delta^2 \Gamma[\psi]}{\delta \psi^2} \psi' \psi'''
\end{equation}
after imposing the vacuum condition.

There are two terms proportional to the inverse propagator, namely $3 \frac{\delta^2 \Gamma[\psi]}{\delta \psi^2} \psi''^2$ and $4 \frac{\delta^2 \Gamma[\psi]}{\delta \psi^2} \psi' \psi'''$, the on-shell condition is only relevant for the second term, to understand this consider the definitions of $\phi^2$ and $\phi^3$
\begin{equation}
    \phi^2 (x) = \int \frac{d^dp_1}{(2\pi)^d} \frac{d^dp_2}{(2\pi)^d} \phi(p_1)\phi(p_2) e^{i (p_1 + p_2)x}
\end{equation}
and
\begin{equation}
    \phi^3 (x) = \int \frac{d^dp_1}{(2\pi)^d} \frac{d^dp_2}{(2\pi)^d} \frac{d^dp_3}{(2\pi)^d}  \phi(p_1)\phi(p_2)\phi(p_3) e^{-i (p_1 + p_2+p_3)x}
\end{equation}
the term $3\frac{\delta^2 \Gamma[\psi]}{\delta \psi^2 }\psi''^2$ picks up factors from $\phi^2$ as indicated by $\psi''$ and the coefficient 3, which indicates the three independent momenta combinations $p_1 + p_2, p_1 + p_3$ and $p_1 + p_4$, thus the term is of the form $\sim(p_i + p_j)^2 - m^2$ which does not vanish on-shell.

On the other hand if we consider the term $4 \frac{\delta^2 \Gamma[\psi]}{\delta \psi^2} \psi' \psi'''$ it either picks up momenta from $\phi$ as indicated by $\psi'$ or from $\phi^3$ through $\psi'''$, with the coefficient 4 indicating that there are 4 independent choices $p_1,p_2,p_3$ and $p_4$, using the fact that $p_1 + p_2 + p_3 = -p_4$ by momentum conservation, it is clear that $\phi^3$ contributes in the same way as $\phi$ in the four point Feynman rule, thus the contribution has the form $\sim p_i^2 - m^2$ which vanishes on-shell with $p_i^2 = m^2$

The connected function is referred to as covariant if under the redefinition $\phi^{a_i} \to \psi^{b_i}(\phi)$ it transforms as 
\begin{equation}
    \mathcal{A}_{a_1 \dots a_n} \to \frac{\delta \psi^{b_1}}{\delta \phi^{a_1}} \frac{\delta \psi^{b_2}}{\delta \phi^{a_2}} \dots \frac{\delta \psi^{b_n}}{\delta \phi^{a_n}} \mathcal{A}_{b_1 \dots b_n}
\end{equation}
with suppressed indices this corresponds to all terms proportional to $\psi'', \psi''' \dots \psi^{(n)}$ vanishing.

\begin{figure}
    \centering
    \begin{tikzpicture}
    \begin{feynman}
        \vertex[blob] (m) at ( 0, 0) {\contour{white}{}};
      \vertex (a) at (-1,-1) ;
      \vertex (b) at ( 1,-1) ;
      \vertex (c) at (-1, 1);
      \vertex (d) at ( 1, 1);
      \vertex[label={right:$=$}] (equals) at (1.25,0);
      \vertex (n) at (3.25,0);
      \vertex (a1) at (2.25,-1);
      \vertex (b1) at (4.25,-1);
      \vertex (c1) at (2.25,1);
      \vertex (d1) at (4.25,1);
      \vertex (n1) at (6,0);
      \vertex (a2) at (7,0);
      \vertex (b2) at (5,1);
      \vertex (c2) at (5,-1);
      \vertex (a3) at (7.5,0);
      \vertex (b3) at (8.5,0);
      \vertex (n2) at (10,0);
      \vertex (a4) at (9,0);
      \vertex (b4) at (11,1);
      \vertex (c3) at (11,-1);
      \vertex[label={right:$+$}] (plus) at (4.5,0);
      \diagram* {
        (a) -- [scalar] (m) -- [scalar] (c),
        (b) -- [scalar](m) -- [scalar] (d),
      };
      \diagram* {
        (a1) -- [scalar] (n) -- [scalar] (c1),
        (b1) -- [scalar](n) -- [scalar] (d1),
      };
        \diagram* {
        (a2) -- [scalar] (n1) -- [scalar] (b2),
        (n1) -- [scalar] (c2),
      };
      \diagram* {
        (a3) -- [scalar] (b3),
      };
      \diagram* {
        (a4) -- [scalar] (n2) -- [scalar] (b4),
        (n2) -- [scalar] (c3),
      };
    \end{feynman}
    \end{tikzpicture}
    \caption{Four point connected amputated function as sum of four point 1PI and gluing of two 3 point 1PI diagrams}
    \label{fig:4 point connected function}
\end{figure}
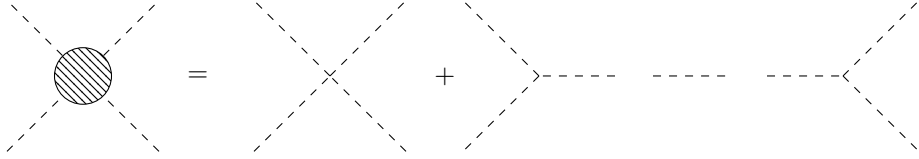

Connected $n$-point functions $\mathcal{A}_{a_1 \dots a_n}$ can be constructed by connecting 1PI functions with $n \geq 3$ to one another using propagators $D_F$ as internal lines, which will be referred to as "gluing" in this paper, and forming all possible tree level topologies as shown in Fig.~\ref{fig:4 point connected function}
The covariance of the connected amputated functions $\mathcal{A}_{a_1 \dots a_n}$ is obscured in this case, it follows from a detailed series of cancellations between the different tree level diagrams, in addition to imposing the on-shell condition for all the external legs.

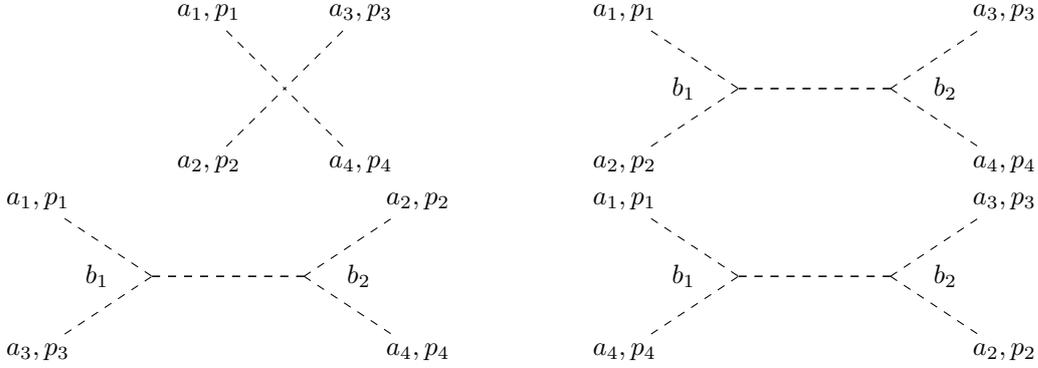
\begin{figure}[h!]
    \centering
    \begin{subfigure}[b]{0.5\textwidth}
        \centering
        \begin{tikzpicture}[baseline = (q.base)]
  \begin{feynman} [inline = (q.base)]
      \vertex (a) {$a_1,p_1$};
      \vertex[below = 1cm of a](q);
      \vertex[right = 1cm of q](p);
      \vertex [below = 2cm of a] (b) {$a_2,p_2$} ;
      \vertex [right = 2cm of a] (c) {$a_3,p_3$} ;
      \vertex [below = 2cm of c] (d) {$a_4,p_4$} ;
      \diagram*{
        (a) -- [scalar] (p) ,
        (b) -- [scalar] (p) ,
        (c) -- [scalar] (p) ,
        (d) -- [scalar] (p) ,
      };
  \end{feynman}
\end{tikzpicture}
    \end{subfigure}%
    ~ 
    \begin{subfigure}[b]{0.5\textwidth}
        \begin{tikzpicture}[baseline = (b.base)]
  \begin{feynman} [inline = (b.base)]
      \vertex (a) {$a_1,p_1$};
      \vertex [below= 1cm of a] (f) ;
      \vertex [right=1.5 cm of f] (b) ;
      \vertex [right= 0.5cm of f,label={[above right= 0.1cm and 1.1cm]:$b_1$}] (c)  ;
      \vertex [below = 2cm of a] (d) {$a_2,p_2$} ;
      \vertex [right = 5cm of a] (h) {$a_3,p_3$} ;
      \vertex [below = 1cm of h] (k) ;
      \vertex [left = 0.5 cm of k,label={[above left= 0.1cm and 1.1cm]:$b_2$}] (q);
      \vertex [left = 1.5 cm of k] (l) ;
      \vertex [below = 2cm of h] (g) {$a_4,p_4$} ;

      \diagram*{
      (a) -- [scalar] (b) --[scalar] (l) -- [scalar] (g),
      (d) -- [scalar] (b)  --[scalar] (l) -- [scalar] (h),
      };
  \end{feynman}
\end{tikzpicture}
    \end{subfigure}
    \begin{subfigure}[b]{0.5\textwidth}
     \begin{tikzpicture}[baseline = (b.base)]
  \begin{feynman} [inline = (b.base)]
      \vertex (a) {$a_1,p_1$};
      \vertex [below= 1cm of a] (f) ;
      \vertex [right=1.5 cm of f] (b) ;
      \vertex [right= 0.5cm of f,label={[above right= 0.1cm and 1.1cm]:$b_1$}]  (c)  ;
      \vertex [below = 2cm of a] (d) {$a_3,p_3$} ;
      \vertex [right = 5cm of a] (h) {$a_2,p_2$} ;
      \vertex [below = 1cm of h] (k) ;
      \vertex [left = 0.5 cm of k,label={[above left= 0.1cm and 1.1cm]:$b_2$}] (q) ;
      \vertex [left = 1.5 cm of k] (l) ;
      \vertex [below = 2cm of h] (g) {$a_4,p_4$} ;

      \diagram*{
      (a) -- [scalar] (b) --[scalar] (l) -- [scalar] (g),
      (d) -- [scalar] (b)  --[scalar] (l) -- [scalar] (h),
      };
  \end{feynman}
\end{tikzpicture}
    \end{subfigure}%
    ~ 
    \begin{subfigure}[b]{0.5\textwidth}
        \begin{tikzpicture}[baseline = (b.base)]
  \begin{feynman} [inline = (b.base)]
      \vertex (a) {$a_1,p_1$};
      \vertex [below= 1cm of a] (f) ;
      \vertex [right=1.5 cm of f] (b) ;
      \vertex [right= 0.5cm of f,label={[above right= 0.1cm and 1.1cm]:$b_1$}]  (c) ;
      \vertex [below = 2cm of a] (d) {$a_4,p_4$} ;
      \vertex [right = 5cm of a] (h) {$a_3,p_3$} ;
      \vertex [below = 1cm of h] (k) ;
      \vertex [left = 0.5 cm of k,label={[above left= 0.1cm and 1.1cm]:$b_2$}]  (q) ;
      \vertex [left = 1.5 cm of k] (l) ;
      \vertex [below = 2cm of h] (g) {$a_2,p_2$} ;

      \diagram*{
      (a) -- [scalar] (b) --[scalar] (l) -- [scalar] (g),
      (d) -- [scalar] (b)  --[scalar] (l) -- [scalar] (h),
      };
  \end{feynman}
\end{tikzpicture}
    \end{subfigure}
    \caption{Diagrams contributing to four point connected function. Labels correspond to the type of particle at each leg and their respective momenta to indicate the different channels}
    \label{fig:four point with labels}
\end{figure}

To identify the necessary cancellations between the different diagrams, it is crucial to count the number of topologies correctly. At the tree level, the number of channels is equal to $n!/|\text{Aut}|$ where $n$ is the number of external legs and $|\text{Aut}|$ is the order of the automorphism group of the diagram, which accounts for the symmetries of a given diagram, for example consider the first two diagrams in Fig.~\ref{fig:four point with labels}, in the first diagram swapping any of the legs with each other results in the same diagram, thus $|\text{Aut}| = 4!$ and the number of diagrams $n!/|\text{Aut}|=1$. For the second diagram notice that swapping $a_1 \leftrightarrow a_2$, $a_3 \leftrightarrow a_4$ or $\{a_1,a_2,b_1\} \leftrightarrow \{a_3,a_4,b_2\}$ all leave the diagram unchanged, each swap has order two, thus $|\text{Aut}| = 8$ and thus the number of distinct diagrams is $n!/|\text{Aut}| = 3$, with all three diagrams shown in the figure. Identifying the full set of Feynman diagrams and their symmetries for a given number of external points $n$ is a combinatorial and graph theory problem, whose solution is not known to the best of my knowledge, though several similar sequences are well known and thoroughly studied as mentioned in the following section.
\section{Counting Tree Graphs}
\label{sec:Counting Tree Graphs}
The construction of the connected function from 1PI vertices requires the identification of all the possible tree diagrams as well as the number of topologies associated with each diagram. For $n \geq 4$ external legs there is more than one type of diagram contributing to the connected function, assigning arbitrary labels to the particles, using $a_i$ for external legs and $b_i$ for internal legs one finds

\subsection{Trees as Integer Partitions}
\label{subsec: Trees as Integer Partitions}
\begin{center}

\begin{figure}[h!]
    \centering
    \begin{subfigure}[b]{0.5\textwidth}
    \centering
     \begin{tikzpicture}[baseline = (k.base)]
  \begin{feynman} [inline = (k.base)]
      \vertex (a) ;
      \vertex[below = 0.5cm of a] (b);
      \vertex [below= 1cm of a] (k) ;
      \vertex [right= 1cm of k] (c) ;
      \vertex [below= 1.5cm of a] (d) ;
      \vertex [below= 2cm of a] (e) ;
      \vertex [right=1 cm of c] (f) ;
      \vertex [right= 1cm of f] (g)  ;
      \vertex [above = 1cm of g] (h) ;
      \vertex [below = 1cm of g] (i)  ;

      \diagram*{
      (a) -- [scalar] (c),
      (b) -- [scalar] (c),
      (d) -- [scalar] (c),
      (e) -- [scalar] (c),
      (c) -- [scalar] (f),
      (f) -- [scalar] (h),
      (f) -- [scalar] (i)
      };
  \end{feynman}
\end{tikzpicture}
    \end{subfigure}%
    ~ 
    \begin{subfigure}[b]{0.5\textwidth}
    \centering
        \begin{tikzpicture}[baseline = (k.base)]
  \begin{feynman} [inline = (k.base)]
      \vertex (a) ;
      \vertex[below = 1cm of a] (b);
      \vertex [below= 1cm of a] (k) ;
      \vertex [right= 1cm of k] (c) ;
      \vertex [below= 2cm of a] (e) ;
      \vertex [right=1 cm of c] (f) ;
      \vertex [right= 1cm of f] (g)  ;
      \vertex [above = 1cm of g] (h) ;
      \vertex [below = 1cm of g] (i)  ;

      \diagram*{
      (a) -- [scalar] (c),
      (b) -- [scalar] (c),
      (e) -- [scalar] (c),
      (c) -- [scalar] (f),
      (f) -- [scalar] (g),
      (f) -- [scalar] (h),
      (f) -- [scalar] (i)
      };
  \end{feynman}
\end{tikzpicture}
    \end{subfigure}
    \caption{Diagrams contributing to 6 point connected function with 1 gluing}
\label{fig:6 point 1 gluing}
\end{figure}
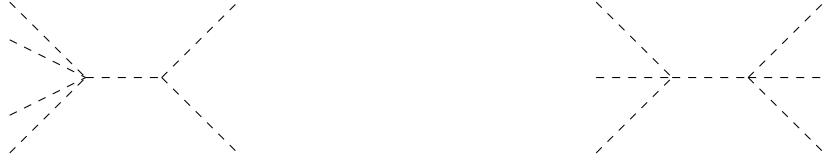   
\end{center}
For $n=6$ external legs, the possibility of two different diagrams with the same number of propagators emerges as shown in ~\ref{fig:6 point 1 gluing}. The identification of how many such diagrams appear for any number of external legs $n$ and number of vertices $k$ can be done using integer partitions. As shown in ~\ref{fig:four point with labels}, the number of labels in a diagram is equal to $n+2k-2$, while the number of vertices in a tree level diagram is equal to $k$.

Let $p_k(n,r)$ denote the number of partitions of $n$ into $k$ parts with minimum size $r$, this function can be written as
\begin{equation}
    p_k(n,r) = p_k(n-k(r-1))
\end{equation}
where $p_k(n)$ is the number of partitions of $n$ into $k$ parts of arbitrary size.

The number of diagrams can then be identified as the number of ways that $n+2k-2$ can be partitioned into $k$ parts of size $\geq 3$.

\begin{equation}
    p_{k}(n+2k-2-k(3-1)) = p_{k}(n+2k-2-2k) = p_{k}(n-2)
\end{equation}

Using the recursion relation for $p_k(n)$ allows for the structure to be studied further
\begin{equation}
    p_{k}(n-2) = p_{k}(n-k-2) + p_{k-1}(n-3)
\end{equation}

The equation states that an $n$ point diagram with $k$ vertices can be obtained either by replacing one of the vertices with one that has an extra leg, which corresponds to the first term on the right hand side, or by gluing a 3 point vertex to an $n-1$ point diagram with $k-1$ vertices, corresponding to the second term. 

For example consider diagrams with 6 external legs
\begin{nalign}
    p_1(4) = 1
    , \quad p_{2}(4) = 2
    , \quad p_{3}(4) = 1
    , \quad p_{4}(4) = 1
    \label{eq: partitions of 6 point}
\end{nalign}
$p_1(4)$ refers to the case with $k=1$ vertices, which is the contact vertex. For $k=2$ there are two cases; $4 = 3+1$ and $4 = 2+2$ corresponding to the gluing of a five point to a three point and the gluing of two four points respectively as shown in Fig.
~\ref{fig:6 point 1 gluing}. $k=3$ corresponds to the the partition $4 = 2 + 1 +1$ representing the gluing of a four point vertex with two three point vertices. Lastly $k=4$ corresponds to the case $4 = 1+1+1+1$ which is the gluing of four three point vertices.
\subsection{From Feynman Diagrams to Tree Graphs}
\label{subsec: Feynman diagrams to graphs}
\begin{figure}[H]
\begin{center}
\centering
     \begin{subfigure}[b]{0.4\textwidth}
    \centering
\begin{tikzpicture}[baseline = (q.base)]
  \begin{feynman} [inline = (q.base)]
      \vertex (a) {$1$};
      \vertex[below = 1cm of a](q);
      \vertex[right = 1.5cm of q](p);
      \vertex[above = 0.2cm of p](g) ;
      \vertex [below = 2cm of a] (b) {$2$} ;
      \vertex [right = 3cm of a] (c) {$3$} ;
      \vertex [below = 2cm of c] (d) {$4$} ;
      \diagram*{
        (a) -- [scalar] (p) ,
        (b) -- [scalar] (p) ,
        (c) -- [scalar] (p) ,
        (d) -- [scalar] (p) ,
      };
  \end{feynman}
\end{tikzpicture}
    \end{subfigure}
    ~ 
    \begin{subfigure}[b]{0.4\textwidth}
    \centering
\begin{tikzpicture}[baseline = (q.base)]
  \begin{feynman} [inline = (q.base)]
      \vertex (a) {$1$};
      \vertex[below = 1cm of a](q);
      \vertex[right = 1.5cm of q](p);
      \vertex[above = 0.2cm of p](g) {$5$} ;
      \vertex [below = 2cm of a] (b) {$2$} ;
      \vertex [right = 3cm of a] (c) {$3$} ;
      \vertex [below = 2cm of c] (d) {$4$} ;
      \diagram*{
        (a) -- [scalar] (p) ,
        (b) -- [scalar] (p) ,
        (c) -- [scalar] (p) ,
        (d) -- [scalar] (p) ,
      };
  \end{feynman}
\end{tikzpicture}
    \end{subfigure}
    \caption{Labeling of Feynman diagrams (left) compared to tree graphs (right)}
    \label{fig:Feynman to Tree Graph}
\end{center}
\end{figure}
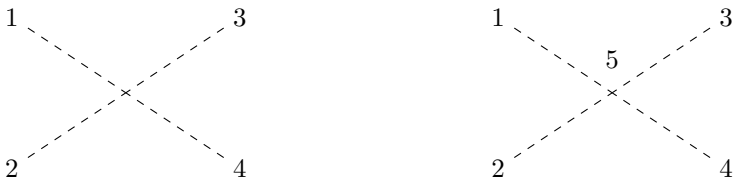
To study the symmetries of tree level Feynman diagrams it is helpful to convert them into tree graphs, so that methods from graph theory may be applied. In the language of graph theory, a point in the graph is referred to as a vertex or a node, while a line is referred to as an edge. A tree is any graph with $m+1$ nodes and $m$ edges. A Feynman diagram with $n$ external legs and $k$ vertices has $n+k$ nodes and $n+k-1$ edges.

A graph is constructed by connecting nodes to one another using edges. To convert a Feynman diagram to a tree graph it is sufficient to add a node whenever two lines cross each other, as well as adding a node at the end of each external line as in ~\ref{fig:Feynman to Tree Graph}.

\begin{figure}[H]
\begin{center}
\centering
    \begin{subfigure}[b]{1\textwidth}
    \centering
     \begin{tikzpicture}[baseline = (b.base)]
  \begin{feynman} [inline = (b.base)]
      \vertex (a) {$1$};
      \vertex [below= 1cm of a] (f) ;
      \vertex [right=1.5 cm of f] (b) ;
      \vertex [right= 0.5cm of f] (c) {$3$} ;
      \vertex [below = 2cm of a] (d) {$2$} ;
      \vertex [right = 5cm of a] (h) {$5$} ;
      \vertex [below = 1cm of h] (k) ;
      \vertex [left = 0.5 cm of k] (q) {$4$};
      \vertex [left = 1.5 cm of k] (l) ;
      \vertex [below = 2cm of h] (g) {$6$} ;

      \diagram*{
      (a) -- [scalar] (b) --[scalar] (l) -- [scalar] (g),
      (d) -- [scalar] (b)  --[scalar] (l) -- [scalar] (h),
      };
  \end{feynman}
\end{tikzpicture}
    \end{subfigure}
    \begin{subfigure}[b]{0.4\textwidth}
    \centering
\begin{tikzpicture}[baseline = (q.base)]
  \begin{feynman} [inline = (q.base)]
      \vertex (a) {$1$};
      \vertex[below = 1cm of a](q);
      \vertex[right = 1.5cm of q](p);
      \vertex[above = 0.2cm of p](g) {$4$};
      \vertex [below = 2cm of a] (b) {$2$} ;
      \vertex [right = 1.5cm of p] (c) {$3$} ;
      
      \diagram*{
        (a) -- [scalar] (p) ,
        (b) -- [scalar] (p) ,
        (c) -- [scalar] (p) ,
      };
  \end{feynman}
\end{tikzpicture}
    \end{subfigure}
    \begin{subfigure}[b]{0.4\textwidth}
    \centering
\begin{tikzpicture}[baseline = (q.base)]
  \begin{feynman} [inline = (q.base)]
      \vertex (a) {$1$};
      \vertex[below = 1cm of a](q);
      \vertex[right = 1.5cm of q](p);
      \vertex[above = 0.2cm of p](g) {$4$};
      \vertex [below = 2cm of a] (b) {$2$} ;
      \vertex [right = 1.5cm of p] (c) {$3$} ;
      
      \diagram*{
        (a) -- [scalar] (p) ,
        (b) -- [scalar] (p) ,
        (c) -- [scalar] (p) ,
      };
  \end{feynman}
\end{tikzpicture}
    \end{subfigure}
    
    \caption{Tree graph and its sub-trees}
    \label{fig: tree and subtrees}
\end{center}
\end{figure}
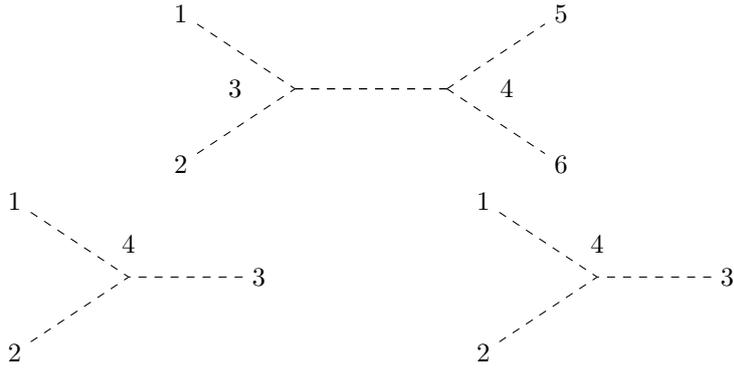

A tree diagram with gluing can be treated as a tree graph composed of two or more sub-trees, with the propagators being treated as lines connecting the central nodes to one another, thus the number of nodes when gluing two sub-trees to one another is 2 less than the sum of nodes of the individual sub-trees as in ~\ref{fig: tree and subtrees}.

Since propagators do not connect to each other, but rather only to vertices in Feynman diagrams, the case of the internal edge being divided into two edges, i.e having a node in the middle of it, never arises. However, it is trivial to see that such a node would not change the symmetries of the diagram and thus it could be removed in any case.\\

\subsection{Counting topologies}
\label{subsec:Counting topologies}
The degree of a node refers to the number of edges to which it is connected. A node of degree 1 is called a leaf or an external node, which corresponds to an external leg in a Feynman diagram. Trees in which all nodes have degree 1 or degree $\geq 3$ are special, they are referred to as \textit{series reduced trees} or as phylogenetic trees when considering rooted\footnote{A root is simply a specification of an external node as a starting point in a tree graph} trees.\cite{ElveyPrice2020:PhylogeneticTrees}. 

The set of tree level Feynman diagrams is precisely that of series reduced trees, which can be proven by induction; starting with four nodes for the base case, it is clear that there is only one tree graph, which is the one shown in the second line of Fig.~\ref{fig: tree and subtrees} and similarly there is only one Feynman diagram. Feynman diagrams are unrooted trees, a Feynamn diagram with $n$ external legs and $k$ vertices corresponds to a phylogenetic tree of type $(n-2,k)$\footnote{The convention for Ward numbers states that an $(n-2,k)$ phylogenetic tree is a rooted tree with $n-1$ external nodes and $k$ internal nodes}, the Ward numbers \cite{Ward1934:PhylogeneticTrees} counting such trees are generated by the recursion relation
\begin{equation}
    W(n-2,k) = kW(n-3,k) + (n+k-3)W(n-3,k-1)
\end{equation}
meaning that a tree of type $(n-2,k)$ can be obtained either by attaching an edge to an internal node in an $(n-3,k)$ type tree or by splitting an edge in two and then attaching an edge to the new node in an $(n-3,k-1)$ type tree. The first case corresponds to replacing a 1PI vertex with one that has one extra leg, while the second case corresponds to attaching a 3 point vertex to a diagram with one less gluing, thus proving the equivalence of Feynman diagrams and phylogenetic or series reduced trees.

The total number of rooted series reduced diagrams corresponds to the sum over Ward numbers for a fixed number of external nodes $n$, it can be found in the Online Enyclopedia of Integer Sequences as \textbf{OEIS A000311} \cite{oeis}, which reads 

\begin{table}[h!]
    \centering
    \begin{tabular}{c|c|c|c|c|c|c}
        $n-2$ &  0 & 1 & 2 & 3 & 4 & 5\\
        \hline
        $W(n-2)$ & 1 & 1 & 4 & 26 & 236 & 2752 \\
    \end{tabular}
    \caption{Number of Feynman diagrams $W(n-2)$ for $n=2,\dots,7$}
    \label{tab:total num of series reduced trees}
\end{table}
Since Feynman diagrams are unrooted, the table counts using $n-2$ such that $W(n-2)$ corresponds to the number of Feynman diagrams with $n$ external legs, starting from $n=2$ as the porpagator.

The Ward numbers defined by the recursion relation are given by \textbf{OEIS A269939} \footnote{The table in the reference starts at n=1 corresponding to a graph with 2 external nodes} \cite{oeis}, adapting the table to the case of Feynman diagrams yields

\begin{table}[H]
    \centering
    \begin{tabular}{ccccccc}
         n-2 \textbackslash k& 1 & 2 & 3 & 4 & 5 &6  \\
         \hline
         1 &  1 \\
         2 & 1 & 3\\
         3 & 1 & 10 & 15\\
         4 & 1 & 25 & 105 & 105\\
         5 & 1 & 56 & 490 & 1260 & 945\\
         6 & 1 & 119 & 1918 & 9450 & 17325 & 10395\\
    \end{tabular}
    \caption{Number of Feynman diagrams with $n$ external legs and $k$ vertices}
    \label{tab: Ward Numbers}
\end{table}
Since a tree of type $(n-2,k)$ corresponds to a rooted series reduced tree with $n-1$ external nodes and $k$ internal nodes, this table is naturally off-set to $n-2$ to match the number of Feynman diagrams with $n$ external legs and $k$ vertices.

As mentioned earlier there are multiple combinations of vertices at the same level of gluing that contribute to the connected function such as in Fig.~\ref{fig:6 point 1 gluing}, which creates a need for going beyond the Ward numbers. A natural refinement can be obtained by examining a multivariate polynomial. The multivariate Ward polynomial is obtained by labeling each vertex by $a_{i-2}$ where $i$ is the number of legs \cite{HAIMAN1989:LagrangeInversion}. For the 6 point case, corresponding to $W(4,k)$ it gives
\begin{nalign}
    & W_{4,1}(a_1,a_2,a_3,a_4) = \frac{5!}{1!}\bigg(\frac{a_4}{5!}\bigg) = a_4\\
& W_{4,2}(a_1,a_2,a_3,a_4) = \frac{6!}{1!1!}\bigg(\frac{a_3}{4!}\bigg)\bigg(\frac{a_1}{2!}\bigg) + \frac{6!}{2!}\bigg(\frac{a_2}{3!}\bigg)^2 = 15a_3 a_1 + 10 a_2^2 \\
& W_{4,3}(a_1,a_2,a_3,a_4) = \frac{7!}{2!1!}\bigg(\frac{a_2}{3!}\bigg)\bigg(\frac{a_1}{2!}\bigg)^2  = 105 a_2 a_1^2\\
& W_{4,4}(a_1,a_2,a_3,a_4) = \frac{8!}{4!1!}\bigg(\frac{a_1}{2!}\bigg)^4  = 105 a_1^4
\end{nalign}
which can be seen as assigning weights to the the partitions identified in (\ref{eq: partitions of 6 point}).

The multivariate Ward polynomial bears close resemblance to the Bell polynomial ~\cite{Comtet_Combinatorics,Riordan_Combinatorial}. In fact upon close examination it is clear that

\begin{equation}
    W_{n,k}(a_1,\dots, a_n) = \frac{(n+k)!}{n!}B_{n,k}(a_1/2,a_2/3,\dots, a_n/(n+1))
    \label{Ward Bell relation}
\end{equation}

The multivariate polynomial is known, its coefficients (up to signs) can be found as \textbf{OEIS Sequence A134685} ~\cite{oeis}, the first few polynomials are given by the table below
\begin{table}[H]
    \centering
    \begin{tabular}{ccccccc}
         n-2 \textbackslash k& 1 & 2 & 3 & 4 & 5  \\
         \hline
         1 &  $a_1$ \\
         2 & $a_2$ & $3a_1^2$\\
         3 & $a_3$ & $10 a_2 a_1$ & $15 a_1^3$\\
         4 & $a_4$ & $10 a_2^2 + 15 a_3 a_1$ & $105 a_2 a_1^2$ & $105 a_1^4$\\
         5 & $a_5$ & $21 a_4 a_1 + 35 a_3 a_2$ & $210 a_3 a_1^2 + 280 a_2^2 a_1$ & $1260 a_2 a_1^3$ & $945 a_1^5$\\
    \end{tabular}
    \caption{Multivariate Ward polynomial for $n$ external nodes and $k$ internal nodes}
    \label{tab: Multivariate Ward Polynomial}
\end{table}
to recover the Ward polynomial simply evaluate $W_{n,k}(1,x,x^2,\dots)$ and it is trivial to see that the right hand side fulfills the recursion relation defining the Ward numbers. The full tree sums given by $W_n$ can be obtained by evaluating $W_{n,k}(1,1,\dots)$. The multivariate Ward polynomial describes the Faà di Bruno algebra \cite{FIGUEROA:HopfAlgebra, Connes2000:Hopf1} and can be related to Feynman graphs ~\cite{Connes2001:Renormalization,Abdesselam2002:Feynman}.

An alternative path towards the derivation of the multivariate Ward polynomial in the context of tree level Feynman diagrams can be done as follows. First consider all trees that are built entirely using cubic vertices for an arbitrary number of propagators $k$. 

The sequence reads
$$ 1, 3,15,105,945,\dots$$
as found in \cite{Bern_2008}, which corresponds to the case $W(n,n)$. If the number of edges is denoted by $m$ and the number of diagrams by $f_n$, then the sequence is given by
\begin{equation}
    f_n = m f_{n-1} = \frac{m!}{(m-1)!}f_{n-1}
\end{equation}
In the case of diagrams made up of four point vertices, they have multiplicities of $1,10,280$, with the sequence continuing to $15400,1401400$ for four and five insertions of the vertex. The sequence is defined by the following recursion
\begin{equation}
    f_n = \frac{m(m+1)}{2} f_{n-1} = \frac{(m+1)!}{2!(m-1)!}f_{n-1}
\end{equation}
\begin{figure}
\centering
\begin{subfigure}[b]{0.25\textwidth}
    \centering
    \begin{tikzpicture}[baseline = (c.base)]
        \begin{feynman} [inline = (c.base)]
            \vertex (c) ;
            \vertex[left = 1cm of c] (b1);
            \vertex[right = 1cm of c] (a3);
            \vertex[above = 1cm of b1] (a1);
            \vertex[below = 1cm of b1] (a2);

            \diagram*{
            (a1) -- [scalar] (c),
            (a2) -- [scalar] (c),
            (a3) -- [scalar] (c)
            };
        \end{feynman}
    \end{tikzpicture}
\end{subfigure}
\begin{subfigure}[b]{0.3\textwidth}
    \centering
    \begin{tikzpicture}[baseline = (c.base)]
        \begin{feynman} [inline = (c.base)]
            \vertex (c) ;
            \vertex[left = 1cm of c] (b1);
            \vertex[right = 1cm of c] (c2);
            \vertex[right = 1cm of c2] (b2);
            \vertex[above = 1cm of b1] (a1);
            \vertex[below = 1cm of b1] (a2);
            \vertex[above = 1cm of b2] (a3);
            \vertex[below = 1cm of b2] (a4);
            \diagram*{
            (a1) -- [scalar] (c),
            (a2) -- [scalar] (c),
            (c) -- [scalar] (c2),
            (a3) -- [scalar] (c2),
            (a4) -- [scalar] (c2),
            };
        \end{feynman}
    \end{tikzpicture}
\end{subfigure}
\begin{subfigure}[b]{0.4\textwidth}
    \centering
    \begin{tikzpicture}[baseline = (c2.base)]
        \begin{feynman} [inline = (c2.base)]
            \vertex (c) ;
            \vertex[left = 1cm of c] (b1);
            \vertex[right = 1cm of c] (c2);
            \vertex[above = 1cm of c2] (a5);
            \vertex[right = 1cm of c2] (c3);
            \vertex[right = 1cm of c3] (b2);
            \vertex[above = 1cm of b1] (a1);
            \vertex[below = 1cm of b1] (a2);
            \vertex[above = 1cm of b2] (a3);
            \vertex[below = 1cm of b2] (a4);
            \diagram*{
            (a1) -- [scalar] (c),
            (a2) -- [scalar] (c),
            (c) -- [scalar] (c2),
            (c2) -- [scalar] (c3),
            (c2) -- [scalar] (a5),
            (a3) -- [scalar] (c3),
            (a4) -- [scalar] (c3),
            };
        \end{feynman}
    \end{tikzpicture}
\end{subfigure}
\par\bigskip
\begin{subfigure}[b]{0.25\textwidth}
    \centering
     \begin{tikzpicture}[baseline = (c.base)]
  \begin{feynman} [inline = (c.base)]
      \vertex (c) ;
      \vertex[left = 1cm of c] (l);
      \vertex[right = 1cm of c] (r);
      \vertex[above = 1cm of r] (d);
      \vertex[below = 1cm of r] (e);
      \vertex[above = 1cm of l] (a);
      \vertex[below = 1cm of l] (b);

      \diagram*{
      (a) -- [scalar] (c),
      (b) -- [scalar] (c),
      (d) -- [scalar] (c),
      (e) -- [scalar] (c)
      };
  \end{feynman}
\end{tikzpicture}
    \end{subfigure}
\begin{subfigure}[b]{0.3\textwidth}
    \centering
        \begin{tikzpicture}[baseline = (c.base)]
  \begin{feynman} [inline = (c.base)]
      \vertex (c1) ;
      \vertex[left = 1cm of c1] (c2);
      \vertex[left = 1cm of c2] (l);
      \vertex[right = 1cm of c1] (r);
      \vertex[above = 1cm of r] (d);
      \vertex[below = 1cm of r] (e);
      \vertex[above = 1cm of l] (a);
      \vertex[below = 1cm of l] (b);

      \diagram*{
      (a) -- [scalar] (c2),
      (b) -- [scalar] (c2),
      (d) -- [scalar] (c1),
      (e) -- [scalar] (c1),
      (l) -- [scalar] (c1),
      (r) -- [scalar] (c2)
      };
  \end{feynman}
\end{tikzpicture}
    \end{subfigure}
\begin{subfigure}[b]{0.4\textwidth}
    \centering
        \begin{tikzpicture}[baseline = (c.base)]
  \begin{feynman} [inline = (c.base)]
      \vertex (c1) ;
      \vertex[left = 1cm of c1] (c2);
      \vertex[left = 1cm of c2] (c3);
      \vertex[left = 1cm of c3] (l);
      \vertex[right = 1cm of c1] (r);
      \vertex[above = 1cm of c2] (f);
      \vertex[below = 1cm of c2] (g);
      \vertex[above = 1cm of r] (d);
      \vertex[below = 1cm of r] (e);
      \vertex[above = 1cm of l] (a);
      \vertex[below = 1cm of l] (b);

      \diagram*{
      (a) -- [scalar] (c3),
      (b) -- [scalar] (c3),
      (d) -- [scalar] (c1),
      (e) -- [scalar] (c1),
      (l) -- [scalar] (c1),
      (r) -- [scalar] (c3),
      (f) -- [scalar] (c2),
      (g) -- [scalar] (c2),
      };
  \end{feynman}
\end{tikzpicture}
    \end{subfigure}
    \caption{Diagrams made of three (upper) and four (lower) point vertices only}
\end{figure}
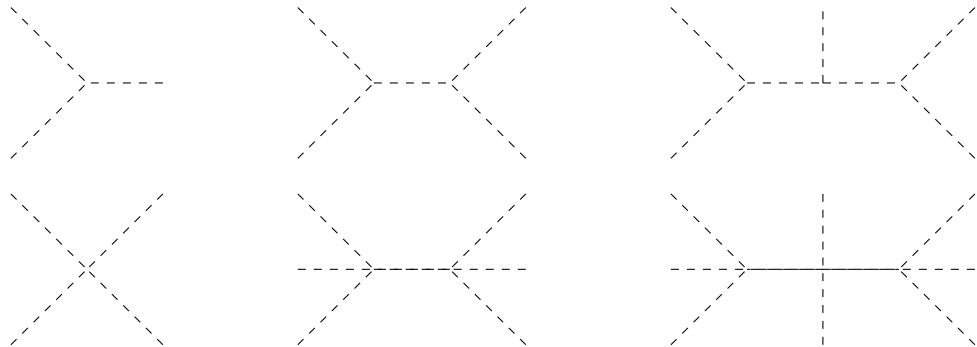 
the pattern continues with diagrams made of only five point diagrams fulfilling

\begin{equation}
    f_n = \frac{m(m+1)(m+2)}{3!} f_{n-1} = \frac{(m+2)!}{3!(m-1)!} f_{n-1}
\end{equation}
and so on.

The proof is quite simple, adding a three point vertex can be treated as adding one more external line to one of the edges in the diagram and there are naturally $m$ choices for where the line could be added. In the case of four point vertices two external lines are added; there are $m$ choices for the first line and $m+1$ choices for the second, but since the order in which the two are added is irrelevant the result must be divided by two to avoid double counting.

The logic continues, namely that if a $k$ point vertex is being glued to the diagram, then it is equivalent to adding $k-2$ external lines, with $m$ options for the first line, $m+1$ for the second up to $m+k-3$ for the $k-2$ line, then to account for the permutations the result must be divided by $(k-2)!$.

Noting that for a diagram given by $W_{n,k}$ the number of edges is $m = n+k-1$, the relations can be used to derive a recursion relation for the polynomial. For example
\begin{nalign}
    W_{4,2}(a_1,a_2,a_3,a_4) &= \binom{5}{3} a_3 W_{1,1}(a_1)+ \binom{5}{2}a_2 W_{2,1}(a_1,a_2)  + \binom{5}{1} a_1 W_{3,1}(a_1,a_2,a_3)\\&  = 10 a_1 a_3 + 10 a_2^2 + 5 a_1 a_3 = 15 a_1 a_3 + 10 a_2^2
\end{nalign}
the recursion relation reads
\begin{equation}
    W_{n,k}(a_1,\dots,a_n) = \sum_{i=1}^{n-k+1} \binom{n+k-1}{i} a_i W_{n-i,k-1}(a_1,\dots,a_{n-i})
\end{equation}
which closely resembles the recursion relation of the Bell polynomials
\begin{equation}
    B_{n+1, k+1}\left(a_1, \ldots, a_{n-k+1}\right)=\sum_{i=0}^{n-k}\binom{n}{i} a_{i+1} B_{n-i, k}\left(a_1, \ldots, a_{n-k-i+1}\right)
\end{equation}
with the two relations agreeing if (\ref{Ward Bell relation}) is substituted into the relation.  
\begin{figure}
    \centering
    \begin{subfigure}[b]{0.5\textwidth}
    \centering
     \begin{tikzpicture}[baseline = (e.base)]
  \begin{feynman} [inline = (e.base)]
      \vertex[label={$a_1$}] (a) ;
      \vertex[below = 1cm of a] (b);
      \vertex[below = 1cm of b, label={$a_2$}] (c);
      \vertex[right = 1cm of b, label={left:$b_1$ \quad}] (d);
      \vertex[right = 1cm of d, label={below:$b_2$}] (e);
      \vertex[right = 1cm of e, label={right:\quad$b_3$}] (f);
      \vertex[right = 1cm of f] (g);
      \vertex[above = 1cm of e, label={right:\quad$b_4$}] (h);
      \vertex[above = 1cm of h] (i);
      \vertex[right = 1cm of i,label={$a_4$}] (j);
      \vertex[left = 1cm of i,label={$a_3$}] (k);
      \vertex[above = 1cm of g,label={$a_5$}] (l);
      \vertex[below = 1cm of g,label={$a_6$}] (m);
      
      \diagram*{
      (a) -- [scalar] (d),
      (c) -- [scalar] (d),
      (d) -- [scalar] (e),
      (e) -- [scalar] (f),
      (e) -- [scalar] (h),
      (h) -- [scalar] (j),
      (h) -- [scalar] (k),
      (f) -- [scalar] (l),
      (f) -- [scalar] (m)
      };
  \end{feynman}
\end{tikzpicture}
    \end{subfigure}%
    ~ 
    \begin{subfigure}[b]{0.5\textwidth}
    \centering
        \begin{tikzpicture}[baseline = (e.base)]
  \begin{feynman} [inline = (e.base)]
      \vertex[label={$a_1$}] (a) ;
      \vertex[below = 1cm of a] (b);
      \vertex[below = 1cm of b,label={$a_2$}] (c);
      \vertex[right = 1cm of b,label={left:$b_1$\quad}] (d);
      \vertex[right = 1cm of d,label={below:$b_2$}] (e);
      \vertex[right = 1cm of e,label={below:$b_3$}] (f);
      \vertex[right = 1cm of f,label={right: \quad $b_4$}] (g);
      \vertex[right = 1cm of g] (h);
      \vertex[above = 1cm of e,label={$a_3$}] (j);
      \vertex[above = 1cm of f,label={$a_4$}] (k);
      \vertex[above = 1cm of h,label={$a_5$}] (l);
      \vertex[below = 1cm of h,label={$a_6$}] (m);

      \diagram*{
      (a) -- [scalar] (d),
      (c) -- [scalar] (d),
      (d) -- [scalar] (e),
      (e) -- [scalar] (f),
      (e) -- [scalar] (j),
      (f) -- [scalar] (g),
      (f) -- [scalar] (k),
      (g) -- [scalar] (l),
      (g) -- [scalar] (m)
      };
  \end{feynman}
\end{tikzpicture}
    \end{subfigure}
    \caption{Two different ways of connecting four 3 point vertices}
    \label{fig: Non-isomorphic 6 point diagrams three gluings}
\end{figure}
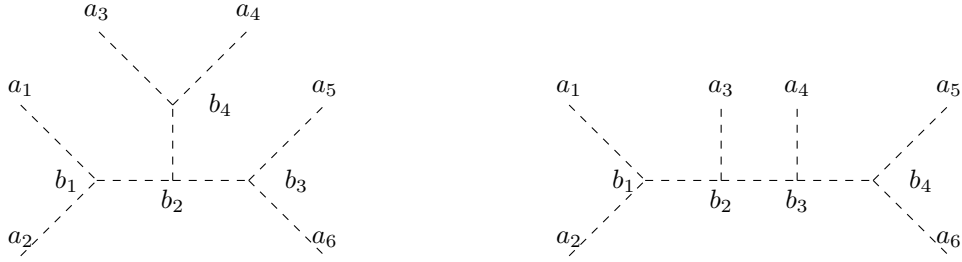

However this is not sufficient to properly count individual diagrams. Even once all the vertices at a certain number of gluings have been identified it is still possible to connect the sub-trees to each other in different ways to obtain unique diagrams.

The graphs in ~\ref{fig: Non-isomorphic 6 point diagrams three gluings} are not isomorphic to one another. A graph isomorphism is a map from the node set of a graph $G$ to the node set of a graph $H$ $\sigma:V(G)\to V(H)$ such that for any pair of nodes $(u,v)$ that form an edge, the nodes $(\sigma(u),\sigma(v))$ also form an edge. In the first diagram the node $b_2$ connects to three other internal nodes $b_1, b_3, b_4$ while in the second diagram it connects to $b_1,b_3,a_3$, thus the map requires mapping $b_4 \to a_3$, but these two nodes form an edge in the first diagram, while not forming one in the second, so there is no isomorphism between the two diagrams.

\subsection{Refined Counting Function}
\label{subsec: Refined counting}
The exact counting function can be identified through the use of generating functionals. To that end note that polynomials can be used to count by assigning a power and numerical coefficient to each possible outcome up to some $n$
\begin{equation}
   P_n(x) = \alpha_0 +\alpha_1 x+ \alpha_2 x^2+ \alpha_3x^3+\dots +x^n = \sum_{i=0}^n \alpha_ix^i
\end{equation}
for example a six sided die can be represented by the polynomial
\begin{equation}
    P_6 =x+x^2+x^3 +x^4 +x^5 +x^6
\end{equation}
then the number of rolls giving a certain outcome when rolling two dice can be identified by squaring $P_6$ and reading off the coefficients.

 In the case of counting the total number of trees I let each power of $x$ correspond to that number of external nodes starting from 1. In terms of generating functionals, the sequences examined earlier are expressed as follows

\begin{equation}
     W(n-2) =   \frac{1}{(n-2)!} \frac{\partial^{n-2}}{\partial x^{n-2}}\sum_{k=1}^{n-2} \frac{(n+k-2)!}{k!} \prod_{j=1}^k \sum_{i=1}^{n-k-1} \frac{x^i}{(i+1)!} \bigg|_{x=0}
\end{equation}

Where $k$ is the number of internal nodes, the factor $1/k!$ accounts for double counting as the internal labels are interchangeable, equivalently seen as taking $k$ products of the same term, due to index contractions, while $(n+k-2)!$ accounts for each node in the graph having a unique label. Lastly the factor $(i+1)!$ accounts for the symmetry of nodes within each vertex after one has been chosen as internal.

Adding a second variable $t$ refines the sequence, now distinguishing the number of diagrams based on the number of vertices $k$, in addition to the number of external legs $n$ \footnote{Corresponds to $k$ internal nodes and $n+1$ external nodes in a rooted tree graph}, which is equivalent to table~\ref{tab: Ward Numbers}

\begin{equation}
   W_{n-2,k} = \frac{1}{(n-2)!} \frac{\partial^{n-2}}{\partial x^{n-2}}  \sum_{k=1}^{n-2}  \frac{1}{k!} \frac{\partial^k}{\partial t^k}\frac{(n+k-2)!}{k!} \prod_{j=1}^k \sum_{i=1}^{n-k-1} \frac{x^i}{(i+1)!}t \bigg|_{x=0,t=0} 
\end{equation}

Assigning weights $a_i$ to each power $x^i$ refines the sequence further to account for the different vertices used to construct a diagram, which yields the multivariate Ward polynomial found in table ~\ref{tab: Multivariate Ward Polynomial}
\begin{equation}
 W_{n-2,k} (a_1,a_2,\dots a_{n-k-1}) = \frac{\partial^{n-2}}{\partial x^{n-2}} \sum_{k=1}^{n-2} \frac{1}{k!} \frac{\partial^k}{\partial t^k}\binom{n+k-2}{k} \bigg(\prod_{j=1}^k \sum_{i=1}^{n-k-1} \frac{a_i x^i}{(i+1)!}\bigg)  t^k \bigg|_{x=0}
\end{equation}

The weights correspond to different vertices, with $a_1$ corresponding to a three point vertex, $a_2$ a four point and so on.

For the case of Feynman diagrams an additional refinement is required. In particular the number of off-shell legs per vertex must be identified.
The refinement is achieved through a new set of weights, since this refinement is specifically for Feynman diagrams, it is reasonable to redefine variables to remove the off-set as follows.

Consider a set of variables $\vec{a}_i = (a_{i,0}, \dots,a_{i,i})$ with $a_{i,0}$ corresponding to a vertex with $i$ legs and all being on-shell, while $a_{i,i}$ refers to the same vertex but with all legs off-shell. Since the focus of this counting is specific to Feynman diagrams, it makes sense to remove the off-set present in the graph theory conventions, thus $\vec{a}_1$ will now be identified with the 1 point vertex, $\vec{a}_2$ the two point and so on, thus the variables $a_i$ will correspond to vertices with $i$ legs, unlike in \ref{subsec:Counting topologies}, where they corresponded to vertices with $i+2$ legs.

The number of Feynman diagrams $\mathcal{F}_{n,k}$ where $n$ is the number of external legs and $k$ the number of vertices is given by the following series

\begin{nalign}
& \sum_{k=1}^{n-2} \mathcal{F}_{n,k} (\vec{a}_1,\vec{a}_2,\dots \vec{a}_{n-k+1}) x^{n-2} t^{2k-2} =\\&a_{n,0} x^{n-2}  + \sum_{k=2}^{n-2}  \frac{n!(k-2)!}{k!} \bigg( \prod_{j=1}^{k} \sum_{i=1}^{n-k-1} \sum_{l=1}^{i+2} \frac{a_{i+2,l}  x^i}{(i+2-l)!(l-1)!} t^{l}  \bigg) 
\label{Refined Counting Function}
\end{nalign}

Within each vertex, the internal legs may be swapped freely with one another and the external legs may be swapped freely with one another as well. A normalization by the factor $k!$ is needed, as there is a product of $k$ identical terms in the generating function and lastly the distinct labels of the external legs and propagators account for the remaining normalization factors. The first term is separated as it corresponds to the contact diagram, which is the only diagram with no gluings at each order and thus possessing trivial symmetries.

The coefficients of this function $\mathcal{F}_{n,k}$ are equal to $n!/|\text{Aut}|$ where $|\text{Aut}|$ is the order of the automorphism group of the Feynman diagram and $n!$ is the number of external legs. Returning to Fig.~\ref{fig: Non-isomorphic 6 point diagrams three gluings}, in the first diagram is invariant under the swaps $a_1 \leftrightarrow a_2$, $a_3 \leftrightarrow a_4$, $a_5 \leftrightarrow a_6$ and $\{a_1,b_1,a_2\} \leftrightarrow \{a_3, b_2, a_4\} \leftrightarrow \{a_5, b_3, a_6 \}$, thus $|\text{Aut}| = 2 \cdot 2 \cdot 2 \cdot 3! = 48$ and the number of distinct topologies is $6!/|\text{Aut}| = 15$. In the second diagram the following swaps leave the diagram invariant $a_1 \leftrightarrow a_2$, $a_5 \leftrightarrow a_6$ and $\{a_1,a_2,b_1,b_2,a_3\} \leftrightarrow\{a_5,a_6,b_4,b_3,a_4\}$ yielding $|\text{Aut}| = 2 \cdot 2 \cdot 2 = 8$ and $6!/|\text{Aut}| = 90$ distinct topologies.

The function is new to my knowledge, its first entries $\mathcal{F}_{n,k}$ read

\begin{table}[h!]
    \centering
    \begin{tabular}{ccccccc}
         n \textbackslash k& 1 & 2 & 3 & 4  \\
         \hline
         2& $a_{2,0}$\\
         3 &  $a_{3,0}$ \\
         4 & $a_{4,0}$ & $3a_{3,1}^2$\\
         5 & $a_{5,0}$ & $10 a_{4,1} a_{3,1}$ & $15 a_{3,2}a_{3,1}^2 $\\
         6 & $a_{6,0}$ & $10 a_{4,1}^2 + 15 a_{5,1} a_{3,1}$ & $45 a_{4,2} a_{3,1}^2 + 60 a_{4,1} a_{3,2} a_{3,1}$ & $90 a_{3,2}^2a_{3,1}^2 + 15 a_{3,3} a_{3,1}^3$\\
         
    \end{tabular}
    \caption{Number of Feynman diagrams for $n$ external legs and $k$ propagators}
    \label{tab: Refined Multivariate Ward Polynomial}
\end{table}
the coefficients are a refinement of the multivariate Ward polynomials found in table \ref{tab: Multivariate Ward Polynomial}

\section{On-Shell Connected Functions and Covariant Feynman Rules}
\label{sec: On-Shell Connected Functions and Covariant Feynman Rules}
The amputated connected function $\mathcal{A}_{n}$ is obtained by taking the sum over the different diagrams contributing to it, which are counted by (\ref{Refined Counting Function}). 
\begin{equation}
    \mathcal{A}_n = \frac{1}{n!} \sum_{S_n} \sum_{k=1}^{n-2} \Delta^{1-k} \mathcal{F}_{n,k}(\vec{a}_1, \vec{a}_2, \dots, \vec{a}_{n-k+1})
    \label{Amputated Connected Function Non-Covariant}
\end{equation}
where the sum over $S_n$ is a sum over all permutations of the external legs.
It remains to show that replacing the coefficients $a_{i+2,l}$ with the individual Feynman rules $\delta^{i+2} \Gamma/ \delta \phi^{i+2}$, applying the vacuum condition to all legs and the on-shell condition to the $i+2-l$ external legs leads to a series cancellations proving that the transformation $\mathcal{A}_n \to \mathcal{A}'_n$ under a field redefinition $\phi \to \psi(\phi)$ is given by
\begin{equation}
\mathcal{A}_n = \left(\frac{\delta \psi}{\delta \phi}\right)^n \mathcal{A}'_n
\end{equation}\\
To simplify notation the variables $z_1,\dots z_n$ will be used, they are defined as 
\begin{equation}
    z_1 = \psi^{\prime}(\phi) \quad ,  z_2 =\psi^{\prime \prime}(\phi) \quad,\dots \quad, z_{n-k+1} = \psi^{(n-k+1)}(\phi)
\end{equation}
accordingly the covariance condition now reads
\begin{equation}
    \mathcal{A}_n = z_1^n \mathcal{A}'_n
\end{equation}\\
the vacuum condition
\begin{equation}
    \frac{\delta}{\delta \phi} \Gamma[\phi]  \bigg|_{\text{vac}} = 0, \quad \frac{\delta}{\delta \psi} \Gamma[\psi]  \bigg|_{\text{vac}} = 0
\end{equation}
is always imposed and the on-shell condition
\begin{equation}
    \frac{\delta^2}{\delta \phi^2} \Gamma[\phi]  \bigg|_{\text{vac},\text{on-shell}} = 0, \quad \frac{\delta^2}{\delta \psi^2} \Gamma[\psi]  \bigg|_{\text{vac},\text{on-shell}} = 0
\end{equation}
will be imposed on the external legs. Additionally it will be sufficient to consider a single permutation of the indices, so the sum over $S_n$ will be suppressed

With the conditions imposed each $n$ point contact vertex contributes 
\begin{equation}
    \frac{\delta^n}{\delta \phi^n} \Gamma[\psi[\phi]] \equiv \sum_{k=1}^n  \frac{\delta^k}{\delta \psi^k} \Gamma(\psi(\phi)) B_{n, k}\left(z_1,z_2,\dots z_{n-k+1})\right) -  \frac{\delta}{\delta \psi}\Gamma[\psi(\phi)] z_n - (n-l)\frac{\delta^2}{\delta \psi^2}\Gamma[\psi(\phi)] z_1z_{n-1}
\end{equation}
where $l$ is the number of legs off-shell.\\

Returning to (\ref{Amputated Connected Function Non-Covariant}) the variables $a_{i+2,l}$ can be identified as

\begin{nalign}
    a_{i+2,l} &= \sum_{r=1}^{i+2} \frac{\delta^r}{\delta \psi^r} \Gamma[\psi(\phi)] B_{i+2,r}(z_1,\dots z_{i+2}) - \frac{\delta}{\delta \psi}\Gamma[\psi(\phi)] z_{i+2} - (i+1-l)\frac{\delta^2}{\delta \psi^2}\Gamma[\psi(\phi)] z_1z_{i+1}\\ & = \tilde{a}_{i+1}  + l b_{i+1} z_1
\end{nalign}
where the variables
\begin{equation}
    \tilde{a}_{i+1} = \sum_{r=1}^{i+2} \frac{\delta^r}{\delta \psi^r} \Gamma[\psi(\phi)] B_{i+2,r}(z_1,\dots z_{i+2}) - \frac{\delta}{\delta \psi}\Gamma[\psi(\phi)] z_{i+2} - (i+1)\frac{\delta^2}{\delta \psi^2}\Gamma[\psi(\phi)] z_1z_{i+1}
\end{equation}
and 
\begin{equation}
    b_{i+1} = \frac{\delta^2}{\delta \psi^2}\Gamma[\psi(\phi)] z_{i+1}
\end{equation} were introduced to simplify the expression. In addition to the transformation from the vertices, the propagators in (\ref{Amputated Connected Function Non-Covariant}) contribute a factor of $1/(\frac{\delta^2}{\delta \psi^2}\Gamma[\psi(\phi)]z_1^{2})^{k-1}$ when transforming. Additionally a factor of $(-1)^{k-1}$ emerges, since each Feynman rule and propagator contributes a factor of $i$, which has been extracted explicitly.

The variable $a_{n,0}$ is defined differently. The contact vertex has all legs on-shell, therefore its transformation behaviors are given by
\begin{nalign}
    a_{n,0} = &\sum_{r=1}^{n} \frac{\delta^r}{\delta \psi^r} \Gamma[\psi(\phi)] B_{n,r}(z_1,\dots z_{n}) - \frac{\delta}{\delta \psi}\Gamma[\psi(\phi)] z_{n} - n\frac{\delta^2}{\delta \psi^2}\Gamma[\psi(\phi)] z_1z_{n-1} \\&= \tilde{a}_{n-1} - b_{n-1} z_1 
\end{nalign}

The sum over all transformation behaviors coming from gluing 1PI vertices to form $n$ point diagrams is then given by

\begin{nalign}
    \frac{\delta^n} {\delta \phi^n} \mathcal{A}_n [\psi(\phi)] =&(\tilde{a}_{n-1}-b_{n-2} z_1)  + \frac{\partial^{n-2}}{\partial x^{n-2}} \sum_{k=2}^{n-2} \frac{n!(k-2)!}{k!(n-2)!(2k-2)!} \frac{(-1)^{k-1}}{z_1^{2k-2}\frac{\delta^2}{\delta \psi^2}\Gamma[\psi(\phi)]^{k-1}} \\ &\bigg( \frac{\partial^{2k-2}}{\partial y^{2k-4}}\prod_{j=1}^{k} \sum_{i=1}^{n-k-1} \frac{x^i}{(i+1)!}\sum_{l=0}^{i+1} \binom{i+1}{l} (\tilde{a}_{i+1}  + l z_1 b_{i+1}) y^{l+1}\bigg) \bigg|_{x,y=0}
    \label{Sum over all diagrams before simplification}
\end{nalign}
simplifying the sums and introducing the two functions $f(x,y)$ and $g(x,y)$

\begin{equation}
    f(x,y) = \sum_{i=1}^{n-k-1} \frac{x^i}{(i+1)!}\tilde{a}_{i+1}(1+y)^{i+1}
\end{equation}
\begin{equation}
    g(x,y) = \sum_{i=1}^{n-k-1} \frac{x^i}{i!} b_{i+1}(1+y)^i 
\end{equation}
simplifies the equation to 
\begin{nalign}
    &(\tilde{a}_{n-1}-b_{n-2} z_1)   + \frac{\partial^{n-2}}{\partial x^{n-2}} \sum_{k=2}^{n-2} \frac{n!(k-2)!}{k!(n-2)!(2k-2)!} \frac{(-1)^{k-1}}{z_1^{2k-2}\frac{\delta^2}{\delta \psi^2}\Gamma[\psi(\phi)]^{k-1}}  \frac{\partial^{2k-2}}{\partial y^{2k-2}} y^{k} \bigg(f(x,y) + y z_1 g(x,y) \bigg)^{k} \bigg|_{x,y=0} 
\end{nalign}
to evaluate the derivatives define $G(g(x,y)) = g(x,y)^j$
then once again by Faa di Bruno
\begin{equation}
     \frac{\partial^{m}}{\partial x^{m}}G(g(x,y)) = \sum_{r=0}^m G^{(r)}(g(x,y))B_{m,r}(g',g'',\dots)
     \label{Faa di Bruno g}
\end{equation}
where $g' = \frac{\partial}{\partial x}g$. Taking the limit $x \to 0$ is simple, first notice
\begin{equation}
    \lim_{x\to 0} g(x,y) = \lim_{x\to 0} \sum_{i=1}^{n-k-1} \frac{x^i}{i!}  b_{i+1}(1+y)^i  = 0
\end{equation}
since the sum starts at $i=0$, thus 
\begin{equation}
    \lim_{x \to 0} G^{(r)} = \lim_{x \to 0} \frac{j!}{(j-r)!}g(x,y)^{j-r} = 0 \quad  \forall r \neq j
\end{equation}
in the case of $j=r$ the result becomes
\begin{equation}
    G^{(j)} = \frac{\partial^j}{\partial g^j} g^j = j!
\end{equation}
thus reducing (\ref{Faa di Bruno g}) to
\begin{equation}
    \frac{\partial^{m}}{\partial x^{m}}G(g(x,y))\bigg|_{x=0} = j! B_{m,j}(g',g'',\dots)\bigg|_{x=0}
\end{equation}
now examine $g$ to extract $g',g'' \dots$
\begin{equation}
    g(x,y) = \sum_{i=1}^{n-k-1} \frac{x^i}{i!}  z_{i+1}(1+y)^i \implies \lim_{x\to0} \frac{\partial^m}{\partial x^m} g(x,y) = \frac{m!}{m!} z_{m+1} (1+y)^m = b_{m+1}(1+y)^m
\end{equation}
thus
\begin{equation}
    \frac{\partial^{m}}{\partial x^{m}}G(g(x,y))\bigg|_{x=0} = j! B_{m,j}(b_2 (1+y),b_3(1+y)^2,\dots)
\end{equation}
the Bell polynomials fulfill the following relation 
\begin{equation}
    B_{n, k}\left(\alpha \beta x_1, \alpha \beta^2 x_2, \ldots, \alpha \beta^{n-k+1} x_{n-k+1}\right)=\alpha^k \beta^n B_{n, k}\left(x_1, x_2, \ldots, x_{n-k+1}\right)
\end{equation}
thus it is possible to extract the $(1+y)$ factors to get
\begin{equation}
    \frac{\partial^{m}}{\partial x^{m}}G(g(x,y))\bigg|_{x=0} = j! (1+y)^m B_{m,j}(b_2,b_3,\dots) =j! (1+y)^m \bigg(\frac{\delta^2}{\delta \psi^2} \Gamma[\psi(\phi)]\bigg)^j B_{m,j}(z_2,z_3,\dots)
\end{equation}
the same steps can be repeated for $f(x,y)$. Evaluating the remaining derivatives and combining the terms leads to

\begin{nalign}
      \frac{\delta^n}{\delta \phi^n} \mathcal{A}_n [\psi(\phi)] = \sum_{k=1}^{n-2}  &\frac{(-1)^{k-1}}{z_1^{2k-2}\frac{\delta^2}{\delta \psi^2}\Gamma[\psi(\phi)]^{k-1}}   \sum_{j=0}^{k} \sum_{m=0}^{n-2} \binom{k-2}{j}   \binom{n-2}{m} j!\frac{(n+k-j-2)!}{(n-2)!} \\& z_1^j \bigg(\frac{\delta^2}{\delta \psi^2} \Gamma[\psi(\phi)]\bigg)^j B_{n-m-2,k-j}(\frac{\tilde{a}_2}{2},\frac{\tilde{a}_3}{3},\dots) B_{m,j}(z_2,z_3,\dots) 
      \label{Unsimplified Tensor Equation}
\end{nalign}

The connected function obtained by summing over all diagram transforms like a tensor if this sum is proportional to powers of $z_1$ only, indeed the claim is that (\ref{Unsimplified Tensor Equation}) simplifies to the following equation
\begin{nalign}
       \frac{\delta^n}{\delta \phi^n} \mathcal{A}_n [\psi(\phi)] & =z_1^{n}\sum_{k=1}^{n-2}   \bigg(\frac{-1}{\frac{\delta^2}{\delta \psi^2}\Gamma[\psi(\phi)]}\bigg)^{k-1} \frac{(n+k-2)!}{(n-2)!}   B_{n-2,k}(\frac{1}{2}\frac{\delta^{3}}{\delta \psi^{3}} \Gamma[\psi(\phi)],\frac{1}{3}\frac{\delta^{4}}{\delta \psi^{4}} \Gamma[\psi(\phi)],\dots) 
      \label{Tensor Equation Final Result}
\end{nalign}

\subsection{Proof of covariance}
\label{subsec: Proof of covariance}
The claim can be proved in a few steps, first take $z_2,z_3,\dots = 0$ to extract the coefficient of $z_1^{n+2}$. 
Notice that the second Bell polynomial depends only on $z_2,\dots$ thus it vanishes except for $m=j=0$. For the first Bell polynomial note that
\begin{nalign}
    &\lim_{z_2,z_3,\dots \to 0} \tilde{a}_{n} = \lim_{z_2,z_3,\dots \to 0} \big(\sum_{r=1}^{n+1} \frac{\delta^r}{\delta \psi^r} \Gamma[\psi(\phi)] B_{n+1,r}(z_1,\dots z_{n+2})\\& - \frac{\delta}{\delta \psi}\Gamma[\psi(\phi)] z_{n+1} - (n+1)\frac{\delta^2}{\delta \psi^2}\Gamma[\psi(\phi)] z_1z_{n} \big) = \frac{\delta^{n+1}}{\delta \psi^{n+1}} \Gamma[\psi(\phi)]z_1^{n+1} 
\end{nalign}
thus the sum becomes
\begin{nalign}
       \frac{\delta^n}{\delta \phi^n} \mathcal{A}_n [\psi(\phi)]&=\sum_{k=1}^{n-2}  \frac{(-1)^{k-1}}{z_1^{2k-2}\frac{\delta^2}{\delta \psi^2}\Gamma[\psi(\phi)]^{k-1}}  \frac{(n+k-2)!}{(n-2)!}  B_{n-2,k}(\frac{\delta^{3}}{\delta \psi^{3}} \Gamma[\psi(\phi)]\frac{z_1^3}{2},\frac{\delta^{4}}{\delta \psi^{4}} \Gamma[\psi(\phi)]\frac{z_1^4}{3},\dots) 
\end{nalign}
extracting powers of $z_1$ gives
\begin{nalign}
      \frac{\delta^n}{\delta \phi^n} \mathcal{A}_n [\psi(\phi)] &=z_1^{n}\sum_{k=1}^{n-2}  \bigg(\frac{-1}{\frac{\delta^2}{\delta \psi^2}\Gamma[\psi(\phi)]}\bigg)^{k-1} \frac{(n+k-2)!}{(n-2)!}   B_{n-2,k}(\frac{1}{2}\frac{\delta^{3}}{\delta \psi^{3}} \Gamma[\psi(\phi)],\frac{1}{3}\frac{\delta^{4}}{\delta \psi^{4}} \Gamma[\psi(\phi)],\dots) 
\end{nalign}
matching equation (\ref{Tensor Equation Final Result}).

It remains to show that other terms vanish, to do so consider terms linear in $z_i$ for any $i \neq 1$, the coefficients of these terms can be extracted from a multivariate Taylor expansion by differentiating with respect to $z_i$ and then taking the limit of all $z_2,z_3, \dots$ to zero, thus keeping only terms linear in $z_i$.

\begin{nalign}
    &\frac{\partial}{\partial z_i}\frac{\delta^n}{\delta \phi^n} \mathcal{A}_n [\psi(\phi)]\bigg|_{z_2,\dots = 0} =\sum_{k=1}^{n-2}  \frac{(-1)^{k-1}}{z_1^{2k-2}}   \sum_{j=0}^{k} \sum_{m=0}^n \binom{k-3}{j}   \binom{n}{m} j!\frac{(n+k-j-2)!}{(n-2)!}  z_1^j \bigg(\frac{\delta^2}{\delta \psi^2} \Gamma[\psi(\phi)]\bigg)^j \\ &\bigg[ \frac{\partial}{\partial z_i } B_{n-m-2,k-j}(\frac{\tilde{a}_2}{2},\frac{\tilde{a}_3}{3},\dots) B_{m,j}(z_2,z_3,\dots) +  B_{n-m-2,k-j}(\frac{\tilde{a}_2}{2},\frac{\tilde{a}_3}{3},\dots) \frac{\partial}{\partial z_i } B_{m,j}(z_2,z_3,\dots) \bigg] \bigg|_{z_2,\dots = 0}
    \label{Differentiation by z_i}
\end{nalign}
For the first term $B_{m,j}(z_2,\dots)$ vanishes except for $m=j=0$ thus the first sum is
\begin{nalign}
    &\sum_{k=1}^{n-2}  \frac{(-1)^{k-1}}{z_1^{2k-2}\frac{\delta^2}{\delta \psi^2}\Gamma[\psi(\phi)]^{k-1}}   \frac{(n+k-2)!}{(n-2)!}  \frac{\partial}{\partial z_i } B_{n-2,k}(\frac{\tilde{a}_2}{2},\frac{\tilde{a}_3}{3},\dots)\bigg|_{z_2,\dots = 0}\\
    & =\sum_{k=1}^{n-2}  \frac{(-1)^{k-1}}{z_1^{2k-2}\frac{\delta^2}{\delta \psi^2}\Gamma[\psi(\phi)]^{k}}   \frac{(n+k-2)!}{(n-2)!} \sum_{r = i}^{n-k}  \frac{\partial \tilde{a}_r}{\partial z_i } \frac{\partial}{\partial \tilde{a}_r }B_{n-2,k}(\frac{\tilde{a}_2}{2},\frac{\tilde{a}_3}{3},\dots)\bigg|_{z_2,\dots = 0}\\
\end{nalign}
for $r+1 \leq i$ the derivative is clearly 0, while for $r+1>i$ 
\begin{equation}
    \frac{\partial \tilde{a}_r}{\partial z_i }= \sum_{l=1}^{r+1} \binom{r+1}{i} \frac{\delta^l}{\delta \psi^l} \Gamma[\psi(\phi)] B_{r-i+1,l-1}(z_1,\dots)
\end{equation}
it is then convenient to split the $r=i$ case from the rest of the sum as follows
\begin{nalign}
    &\sum_{k=1}^{n-2}  \frac{(-1)^{k-1}}{z_1^{2k-2}\frac{\delta^2}{\delta \psi^2}\Gamma[\psi(\phi)]^{k-1}}   \frac{(n+k-2)!}{(n-2)!} z_1 \frac{\delta^2}{\delta \psi^2} \Gamma[\psi(\phi)] \frac{\partial}{\partial \tilde{a}_i }B_{n-2,k}(\frac{\tilde{a}_2}{2},\frac{\tilde{a}_3}{3},\dots)\bigg|_{z_2,\dots = 0}\\
    &+\sum_{k=1}^{n-2}  \frac{(n+k-2)!}{(n-2)! z_1^{2k-2} }\frac{(-1)^{k-1}}{\frac{\delta^2}{\delta \psi^2}\Gamma[\psi(\phi)]^{k-1}} \sum_{r = i+1}^{n-k} \sum_{l=1}^{r+1} \binom{r+1}{i} \frac{\delta^l}{\delta \psi^l} \Gamma[\psi(\phi)] B_{r-i+1,l-1}(z_1,\dots) \frac{\partial}{\partial \tilde{a}_r }B_{n-2,k}(\frac{\tilde{a}_2}{2},\frac{\tilde{a}_3}{3},\dots)\bigg|_{z_2,\dots = 0}\\
\end{nalign}
the derivatives of the Bell polynomials fulfill
\begin{equation}
    \frac{\partial}{\partial \tilde{a}_r }B_{n,k}(\frac{\tilde{a}_2}{2},\frac{\tilde{a}_3}{3},\dots) = \frac{1}{r}\binom{n}{r-1}B_{n-r+1,k-1}(\frac{\tilde{a}_2}{2},\frac{\tilde{a}_3}{3},\dots)
\end{equation}
which following a series of simplifications the first term gives
\begin{nalign}
    &z_1^{n-i} \sum_{k=1}^{n-2}  \frac{(-1)^{k-1}}{\frac{\delta^2}{\delta \psi^2}\Gamma[\psi(\phi)]^{k}}  \binom{n+k-2}{i}  B_{n+k-i-2,k-1}(0,\frac{\delta^{3}}{\delta \psi^{3}} \Gamma[\psi(\phi)],\dots)
    \label{First sum first term}
\end{nalign}
while the second gives
\begin{nalign}
    &z_1^{n-i}\sum_{k=1}^{n-2}   \frac{(-1)^{k-1}}{\frac{\delta^2}{\delta \psi^2}\Gamma[\psi(\phi)]^{k-1}} \binom{n+k-2}{i} \sum_{r = 1}^{n-i-k}  \binom{n+k-i-2}{r} \frac{\delta^{r+2}}{\delta \psi^{r+2}}\Gamma[\psi(\phi)] B_{n+k-i-r-2,k-1}(0,\frac{\delta^{3}}{\delta \psi^{3}} \Gamma[\psi(\phi)],\dots)\\
      &+z_1^{n-i}\sum_{k=1}^{n-2}   \frac{(-1)^{k-1}}{\frac{\delta^2}{\delta \psi^2}\Gamma[\psi(\phi)]^{k-1}} \binom{n+k-2}{i-1}   \sum_{r = 1}^{n-i-k+2}   \binom{n+k-i-1}{r+1}  \frac{\delta^{r+2}}{\delta \psi^{r+2}}\Gamma[\psi(\phi)] B_{n+k-i-r-2,k-1}(0,\frac{\delta^{3}}{\delta \psi^{3}} \Gamma[\psi(\phi)],\dots)
\label{Coefficients to cancel 1}
\end{nalign}
Returning to the second term in (\ref{Differentiation by z_i}) note the identity
\begin{equation}
    \frac{\partial}{\partial z_i } B_{m,j}(z_2,z_3,\dots) = \binom{m}{i-1}B_{m-i+1,j-1}(z_2,z_3,\dots)
\end{equation}
which following a similar set of simplifications as before gives
\begin{nalign}
    z_1^{n-i}\sum_{k=1}^{n-2}   \frac{(-1)^{k-1}}{\frac{\delta^2}{\delta \psi^2}\Gamma[\psi(\phi)]^{k-2}}   (k-2)  \binom{n+k-3}{i-1} B_{n+k-i-2,k-1}(0,\frac{\delta^{3}}{\delta \psi^{3}} \Gamma[\psi(\phi)],\dots)\\
    \label{second sum}
\end{nalign}
combining (\ref{First sum first term}) and (\ref{second sum}) leads to
\begin{nalign}
    &z_1^{n-i} \sum_{k=1}^{n-2}  \frac{(-1)^{k}}{\frac{\delta^2}{\delta \psi^2}\Gamma[\psi(\phi)]^{k-1}}  \binom{n+k-1}{i}  B_{n+k-i-1,k}(0,\frac{\delta^{3}}{\delta \psi^{3}} \Gamma[\psi(\phi)],\dots)\\
    &+ z_1^{n-i}\sum_{k=1}^{n-3}   \frac{(-1)^{k}}{\frac{\delta^2}{\delta \psi^2}\Gamma[\psi(\phi)]^{k-1}}   (k)  \binom{n+k-2}{i-1} B_{n+k-i-1,k}(0,\frac{\delta^{3}}{\delta \psi^{3}} \Gamma[\psi(\phi)],\dots)\\
\end{nalign}
the following recursive Bell polynomial identities can then be used to simplify the sum 
\begin{equation}
    B_{n+1, k+1}\left(x_1, \ldots, x_{n-k+1}\right)=\sum_{i=0}^{n-k}\binom{n}{i} x_{i+1} B_{n-i, k}\left(x_1, \ldots, x_{n-k-i+1}\right)
\end{equation}

\begin{equation}
    k B_{n,k} = \sum^{n-1}_{l=k-1} \binom{n}{l}x_{n-l}B_{l,k-1} = \sum^{n-k+1}_{l=1} \binom{n}{l}x_l B_{n-l,k-1} 
\end{equation}
which yields the following equation
\begin{nalign}
    &z_1^{n-i} \sum_{k=1}^{n-3} \bigg[  \frac{(-1)^{k}}{\frac{\delta^2}{\delta \psi^2}\Gamma[\psi(\phi)]^{k-1}}  \binom{n+k-2}{i} \sum_{r=1}^{n-i-1} \binom{n+k-i-2}{r} \\&\frac{\delta^{r+2}}{\delta \psi^{r+2}}\Gamma[\psi(\phi)] B_{n+k-i-r-2,k-1}(0,\frac{\delta^{3}}{\delta \psi^{3}} \Gamma[\psi(\phi)],\dots)\bigg] \\
     & + z_1^{n-i}\sum_{k=1}^{n-3} \bigg[   \frac{(-1)^{k}}{\frac{\delta^2}{\delta \psi^2}\Gamma[\psi(\phi)]^{k-1}} \binom{n+k-2}{i-1}  \sum_{r=1}^{n-k-1} \binom{n+k-i-1}{r+1} \\& \frac{\delta^{r+2}}{\delta \psi^{r+2}}\Gamma[\psi(\phi)] B_{n+k-i-r-2,k-1}(0,\frac{\delta^{3}}{\delta \psi^{3}} \Gamma[\psi(\phi)],\dots)
     \bigg]
     \label{Coefficients to cancel 2}
\end{nalign}
Note that the summation limit is higher than necessary, since for any $r\geq n-i-1$ the Bell polynomial becomes $B_{k-2,k-1}$ which vanishes since $k-2<k-1$, thus the sum limits for $r$ can be changed to $n-i-1$ freely.
Taking the sum of (\ref{Coefficients to cancel 1}) and (\ref{Coefficients to cancel 2})  it is clear that the sum vanishes thus confirming that linear powers of non-tensor structures vanish.
\begin{equation}
    \frac{\partial}{\partial z_i}\frac{\delta^n}{\delta \phi^n} \mathcal{A}_n [\psi(\phi)]\bigg|_{z_2,\dots = 0} = 0
\end{equation}
Lastly consider higher powers of $z_2,\dots$. As seen from the derivatives of $\tilde{a}_i$ and the partial Bell polynomials $B_{n,k}$ higher derivatives are related by binomials to lower ones, for example

\begin{nalign}
    \frac{\partial}{\partial z_j}  \frac{\partial}{\partial z_i} B_{n,k}(z_2,\dots) &= \frac{\partial}{\partial z_j} \binom{n}{i-1} B_{n-i+1,k-1}(z_2,\dots) \\&= \binom{n}{i-1} \binom{n-i+1}{j-1} B_{n-i-j+2,k-2}(z_2,\dots) 
\end{nalign}
but as shown earlier the Bell polynomials fulfill two binomial recursion relations, thus the cancellation of coefficients of products of $z_i z_j$ and higher powers follows the same steps proving the result and showing that the sum over all tree diagrams results in objects that transform like tensors on-shell.

\subsection{Existence of Covariant Feynman Rules}
\label{subsec: Covariant Feynman rules}
Since the sum over all diagrams on-shell results in a tensor, it is possible to obtain the same results using a simplified form of Feynman rules, in particular if one simply writes down the tensor part of each Feynman rule, then the final result upon summing over all valid diagrams will be the same.

To prove the statement start by considering a new set of variables $\tilde{c}_{i}$ defined by
\begin{equation}
    \tilde{c}_i = \frac{\delta^{i}}{\delta \psi^{i}}\Gamma[\psi(\phi)] z_1^{i}
\end{equation}
the variables encode the transformation behavior under an arbitrary diffeomorphism, which implies in this case that the function in question is simply the tensor part of $\frac{\delta^{i+2}}{\delta \phi^{i+2}}\Gamma[\phi]$, as shown by its transformation behavior. Define the function $\mathcal{B}_n$ as the sum over diagrams with these new variables

\begin{nalign}
  \frac{\delta^n}{\delta \phi^n} \mathcal{B}_n [\psi(\phi)]=&\tilde{c}_{n-1}  + \frac{\partial^{n-2}}{\partial x^{n-2}} \sum_{k=2}^{n-2} \frac{n!(k-2)!}{k!(n-2)!(2k-2)!} \frac{(-1)^{k-1}}{z_1^{2k-2}\frac{\delta^2}{\delta \psi^2}\Gamma[\psi(\phi)]^{k-1}} \\ & \frac{\partial^{2k-2}}{\partial y^{2k-2}}\prod_{j=1}^{k} \sum_{i=1}^{n-k-1} \frac{x^i \tilde{c}_{i+1}}{(i+1)!}\sum_{l=0}^{i+1} \binom{i+1}{l}  y^{l+1} \bigg|_{x,y=0}
    \label{Sum over all diagrams tensor like}
\end{nalign}
in this case there is no dependence on $l$ in the coefficients, thus the simplification proceeds as follows
\begin{nalign}
    &\tilde{c}_{n}  + \frac{\partial^{n-2}}{\partial x^{n-2}} \sum_{k=2}^{n-2} \frac{n!(k-2)!}{k!(n-2)!(2k-2)!} \frac{(-1)^{k-1}}{z_1^{2k-2}\frac{\delta^2}{\delta \psi^2}\Gamma[\psi(\phi)]^{k-1}} \\ & \frac{\partial^{2k-2}}{\partial y^{2k-2}}\bigg( \sum_{i=1}^{n-k-1} \frac{x^i \tilde{c}_{i+1}}{(i+1)!} y(1+y)^{i+1}\bigg)^{k}\bigg|_{x,y=0}
\end{nalign}
which can further be simplified to
\begin{nalign}
    &  \sum_{k=1}^{n-2} \frac{n!(k-2)!}{(n-2)!(2k-2)!} \binom{2k-2}{k}\frac{(-1)^{k-1}}{z_1^{2k-2}\frac{\delta^2}{\delta \psi^2}\Gamma[\psi(\phi)]^{k-1}} \\ & \frac{\partial^{n-2}}{\partial x^{n-2}}\frac{\partial^{k-2}}{\partial y^{k-2}}H(x,y) \bigg|_{x,y=0}
\end{nalign}
where 
\begin{equation}
    H(x,y) = h(x,y)^{k} = \bigg( \sum_{i=1}^{n-k-1} \frac{x^i \tilde{c}_{i+1}}{(i+1)!} y(1+y)^{i+1}\bigg)^{k}
\end{equation}
using Fa\`a di Bruno's formula it is possible to evaluate the derivative as
\begin{equation}
    \frac{\partial^{n-2}}{\partial x^{n-2}} H(x,y) \bigg|_{x=0} = \sum_{r=0}^{n-2} H^{(r)}(h(x,y)) B_{n-2,r}(h',h'',\dots) \bigg|_{x=0} 
\end{equation}
given
\begin{equation}
    \lim_{x \to 0} h(x,y) = 0
\end{equation}
it follows that
\begin{equation}
    \frac{\partial^{n-2}}{\partial x^{n-2}} H(x,y) \bigg|_{x=0} = k! B_{n-2,k}(\frac{\tilde{c}_2}{2}(1+y)^2,\frac{\tilde{c}_3}{3}(1+y)^3,\dots) = (k+1)! (1+y)^{n+k-2}  B_{n-2,k}(\frac{\tilde{c}_2}{2},\frac{\tilde{c}_3}{3},\dots)
\end{equation}
evaluating the derivatives with respect to $y$ then gives a factor of ${(n+k-2)!}/{(n+k-2 - (k-2))!} = {(n+k-2)!}/{n!}$ resulting in
\begin{nalign}
    \frac{\delta^n}{\delta \phi^n} \mathcal{B}_n [\psi(\phi)]=&  \sum_{k=1}^{n-2} \frac{n!(k-2)!}{(n-2)!(2k-2)!} \binom{2k-2}{k}\frac{(-1)^{k-1}}{z_1^{2k-2}\frac{\delta^2}{\delta \psi^2}\Gamma[\psi(\phi)]^{k-1}} \\ & \frac{(n+k-2)!}{n!}k!  B_{n-2,k}(\frac{\tilde{c}_2}{2},\frac{\tilde{c}_3}{3},\dots)
\end{nalign}
simplifying the numerical factors and substituting in the definition of $\tilde{c}_{i}$ leads to
\begin{nalign}
     \frac{\delta^n}{\delta \phi^n} \mathcal{B}_n [\psi(\phi)] = \sum_{k=1}^{n-2} \frac{(n+k-2)!}{(n-2)!} \frac{(-1)^{k-1}}{z_1^{2k-2}\frac{\delta^2}{\delta \psi^2}\Gamma[\psi(\phi)]^{k-1}}  B_{n-2,k}(\frac{1}{2} \frac{\delta^3 \Gamma[\psi(\phi)]}{\delta \psi^3},\frac{1}{3}\frac{\delta^4 \Gamma[\psi(\phi)]}{\delta \psi^4},\dots)
\end{nalign}
which precisely matches (\ref{Tensor Equation Final Result}) thus $\mathcal{B}_n = \mathcal{A}_n$ proving that using a set of covariant building blocks leads to the same final result for the on-shell connected amplitudes.


The formula was derived by assuming that the different index contractions are only relevant for the non-tensor parts of the Feynman rules, this is sufficient for showing covariance and leads to the closed formula for the on-shell amputated function $\mathcal{A}_{a_1\dots a_n}$ provided in~\cite{alminawi2025scalaramps}
\begin{equation}
\mathcal{A}_n=\sum_{k=1}^{n-2}  \frac{(n+k-2)!}{(n-2)!} \Delta^{1-k} B_{n-2, k}\left(\frac{1}{2} \mathcal{R}_3, \ldots, \frac{1}{n-k} \mathcal{R}_{n-k-2}\right)
\end{equation}
, which produces the function up to symmetrization over internal indices, but it is possible to refine the formula further, such that no symmetrization over internal indices is needed.

Denoting the covariant building blocks with $\mathcal{R}_{n} = (\mathcal{R}_{n,0}, \mathcal{R}_{n,1}, \dots, \mathcal{R}_{n,n})$ where the first index is the total number of legs and the second is the number of legs connected to a propagator, the on-shell connected function with suppressed indices is given by
\begin{equation}
    \mathcal{A}_{n} = \frac{1}{n!} \sum_{S_n}\sum_{k=1}^{n-2} \Delta^{1-k}\mathcal{F}_{n,k}(\mathcal{R}_3, \mathcal{R}_4, \dots \mathcal{R}_{n-k}) 
\end{equation}
where $\mathcal{F}_{n,k}(\mathcal{R}_{3},\dots)$ are the polynomials counting Feynman diagrams given by (\ref{Refined Counting Function}) and $\sum_{S_n}$ represents a sum over all the permutations of the indices of $\mathcal{A}_n$. The placement of the indices is straightforward, with the only restriction being that the pair of indices in a propagator may not both be contracted in the same vertex, nor may two vertices be contracted to each other more than once; as that would correspond to a loop. To illustrate the use of the formula $\mathcal{A}_6$ is given by
\begin{align}
\mathcal{A}_{a_1a_2a_3a_4a_5a_6} &= \mathcal{R}_{a_1a_2a_3a_4a_5a_6} 
\\
&
+\frac{1}{6!}\sum_{S_6}\bigg[
15\,
\mathcal{R}_{a_1a_2a_3a_4b_1}\,\Delta^{b_1b_2}(s_{56})\,\mathcal{R}_{b_2a_5a_6}  
\nonumber\\
&+
10\,
\mathcal{R}_{a_1a_2a_3b_1}\,\Delta^{b_1b_2}(s_{123})\,\mathcal{R}_{b_2a_4a_5a_6} 
\nonumber\\
&+
45\,
\mathcal{R}_{a_1a_2b_1}\Delta^{b_1b_2}(s_{12})\,
\mathcal{R}_{a_3a_4b_2b_3}\,\,\Delta^{b_3b_4}(s_{56})\,\mathcal{R}_{b_4a_5a_6} 
\nonumber\\
&
+
60\,
\mathcal{R}_{a_1a_2a_3b_1}\,\Delta^{b_1b_2}(s_{1234})\,\mathcal{R}_{b_2b_3a_4}\,\Delta^{b_3b_4}(s_{56})\,\mathcal{R}_{b_4a_5a_6} 
\nonumber\\
&
+
15\,
\mathcal{R}_{a_1a_2b_1}\,\Delta^{b_1b_2}(s_{12})\,\mathcal{R}_{b_2b_3b_4}\,\Delta^{b_3b_5}(s_{34})\,\mathcal{R}_{b_5a_3a_4}\,
\Delta^{b_4b_6}(s_{56})\,
\mathcal{R}_{b_6a_5a_6} \nonumber
\\
&+ 
90\,
\mathcal{R}_{a_1a_2b_1}\,\Delta^{b_1b_2}(s_{12})\,\mathcal{R}_{b_2b_3a_3}\,\Delta^{b_3b_4}(s_{123})\,\mathcal{R}_{b_4b_5a_4}\,
\Delta^{b_5b_6}(s_{56})\, \mathcal{R}_{b_6a_5a_6} \,\,\,
\bigg]\nonumber
\end{align}
An explicit derivation of the covariant rules is easily achieved through the use of geometry as done in ~\cite{alminawi2025scalaramps,Apostolos2020:Covariance}. However, a derivation that does not rely on geometric methods has not yet been identified and is slated for future works.
\section{Conclusion and Outlook}
\label{sec: Conclusion}
On-shell amputated connected functions for any quantum field theory are covariant under field redefinitions. In this paper, it was shown that the properties follow directly from the Fa\`a di Bruno formula and the definition of the connected function in terms of sums over tree diagrams, as well as the imposition of the vacuum and on-shell conditions. The approach used was fully general, making no assumptions about the type of field or the number of derivatives contained in the definition of the effective action $\Gamma[\phi]$.

The number of tree level diagrams for any number of external legs $n$ was identified in terms of integer partitions and the order of the automorphism groups of the individual diagrams. A general formula for computing these numbers was presented in terms of generating functions. The number of diagrams grows rapidly with $n$, making the task of identifying their symmetries via brute force a computational obstacle that is resolved via the use of the counting formula.

The use of covariant Feynman rules allows for manifestly covariant expressions of the on-shell amputated connected function $\mathcal{A}_n$ to be computed. The validity of using covariant Feynman rules to arbitrary $n$ was verified, which when combined with the counting function allows for a closed form expression for $\mathcal{A}_n$.

The methods used at the tree level lay out a simple path towards extending the results to arbitrary loop order by adjusting the counting function, which is planned for a future work, but it has been explored for 1 and 2 point functions at the 1-loop level in~\cite{Apostolos2020:Covariance,alminawi2025scalaramps}. 

\acknowledgments
The author acknowledges funding
from SNF through the PRIMA grant no. 201508.

I am grateful to Ilaria Brivio and Joe Davighi for collaboration on different works that led to this project, for help with identifying and refining the scope of the motivating problem and for their guidance in the preparation of this manuscript.
\newpage
\bibliographystyle{JHEP}
\bibliography{counting_refs}

\end{document}